\documentclass[5p,times]{elsarticle}

\usepackage{lineno,hyperref}
\usepackage{booktabs}
\usepackage{mathrsfs,amsmath,amsfonts,amssymb}
\usepackage{algorithm}
\usepackage{algorithmicx}
\usepackage{algpseudocode}
\usepackage{subfigure}
\usepackage{blindtext}
\usepackage{enumitem}
\usepackage{color, colortbl}
\usepackage{xcolor}
\usepackage{soul}
\usepackage{tikz}
\usetikzlibrary{tikzmark,calc}
\definecolor{LightCyan}{rgb}{0.68,1,1}
\definecolor{LightPink}{rgb}{0.96,0.76,0.76}
\definecolor{LightBlue}{rgb}{0.63, 0.79, 0.95}

\DeclareRobustCommand{\hlblue}[1]{{\sethlcolor{LightBlue}\hl{#1}}}
\DeclareRobustCommand{\hlpink}[1]{{\sethlcolor{LightPink}\hl{#1}}}

\newcommand\DrawBox[3][]{%
  \begin{tikzpicture}[remember picture,overlay]
    \draw[overlay,fill=gray!30,#1] 
    ([xshift=14em,yshift=1.6ex]{pic cs:#2}) 
    rectangle 
    ([xshift=-26pt,yshift=-0.5ex]pic cs:#3);
  \end{tikzpicture}%
}

\modulolinenumbers[5]

\journal{Journal of \LaTeX\ Templates}

\begin{document}
\begin{sloppypar}

\begin{frontmatter}

\title{Unravelling Token Ecosystem of EOSIO Blockchain}
% \tnotetext[mytitlenote]{Fully documented templates are available in the elsarticle package on \href{http://www.ctan.org/tex-archive/macros/latex/contrib/elsarticle}{CTAN}.}

\author[mymainaddress]{Weilin Zheng}
\ead{zhengwlin@mail2.sysu.edu.cn}

\author[mymainaddress]{Bo Liu}
\ead{liub76@mail2.sysu.edu.cn}

\author[mysecondaryaddress]{Hong-Ning Dai}
\ead{hndai@ieee.org}

\author[mymainaddress]{Zigui Jiang\corref{mycorrespondingauthor}}
\cortext[mycorrespondingauthor]{Corresponding author}
\ead{jiangzg3@mail.sysu.edu.cn}

\author[mymainaddress]{Zibin Zheng}
\ead{zhzibin@mail.sysu.edu.cn}

\author[mythirdaddress]{Muhammad Imran}
\ead{dr.m.imran@ieee.org}

% \author{Elsevier\fnref{myfootnote}}
% \address{Radarweg 29, Amsterdam}
% \fntext[myfootnote]{Since 1880.}

% %% or include affiliations in footnotes:
% \author[mymainaddress,mysecondaryaddress]{Elsevier Inc}
% \ead[url]{www.elsevier.com}

% \author[mysecondaryaddress]{Global Customer Service\corref{mycorrespondingauthor}}
% \cortext[mycorrespondingauthor]{Corresponding author}
% \ead{support@elsevier.com}

\address[mymainaddress]{School of Computer Science and Engineering, Sun Yat-sen University, Guangzhou, China}
\address[mysecondaryaddress]{Department of Computing and Decision Sciences, Lingnan University, Hong Kong, China}
\address[mythirdaddress]{School of Engineering, Information Technology and Physical Sciences, Federation University, Australia}

\begin{abstract}

Being the largest Initial Coin Offering project, EOSIO has attracted great interest in cryptocurrency markets. Despite its popularity and prosperity (e.g., 26,311,585,008 token transactions occurred from June 8, 2018 to Aug. 5, 2020), there is almost no work investigating the EOSIO token ecosystem. To fill this gap, we are the first to conduct a systematic investigation on the EOSIO token ecosystem by conducting a comprehensive graph analysis on the entire on-chain EOSIO data (nearly 135 million blocks). We construct token creator graphs, token-contract creator graphs, token holder graphs, and token transfer graphs to characterize token creators, holders, and transfer activities. 
Through graph analysis, we have obtained many insightful findings and observed some abnormal trading patterns. Moreover, we propose a fake-token detection algorithm to identify tokens generated by fake users or fake transactions and analyze their corresponding manipulation behaviors. Evaluation results also demonstrate the effectiveness of our algorithm.
\end{abstract}

\begin{keyword}
Blockchain, EOSIO, Token, Fake Detection, Graph Analysis
% \texttt{elsarticle.cls}\sep \LaTeX\sep Elsevier \sep template
% \MSC[2010] 00-01\sep  99-00
\end{keyword}

\end{frontmatter}

% \linenumbers

\section{Introduction}
Cryptocurrencies such as~Bitcoin~\cite{nakamoto2019bitcoin} and Ethereum~\cite{wood2014ethereum} have received great interest from investors~\cite{crosby2016blockchain}. As an underlying technology, blockchain has essentially established a distributed database with characteristics like traceability, security, and immutability~\cite{zheng2017overview}. Meanwhile, smart contracts running on top of blockchains can automate business processes, simplify trading actions, and reduce administrative costs~\cite{ZHENG2020475,viriyasitavat2018blockchain,ganne2018can,schmidt2019blockchain}. However, blockchains like Bitcoin and Ethereum suffer from a low transaction throughput due to inefficient consensus protocols~\cite{bach2018comparative,lepore2020survey,cao2020performance}, like Proof-of-Work (PoW). Thus, they are incapable of supporting real-time trading services. 

In contrast to PoW-based blockchain systems~\cite{meneghetti2020survey}, EOSIO adopts a more efficient consensus protocol -- Delegated Proof-of-Stake (DPoS)~\cite{larimer2014delegated,io2017eosdpos}. It allows EOSIO to achieve much higher transaction throughput (up to 8,000 transactions per second) and much lower confirmation latency (within one second) than Bitcoin and Ethereum~\cite{io2018eos}. Consequently, EOSIO has become an attractive option for many decentralized applications (DApps), especially for applications having a stringent requirement on trading time. According to \textit{Crowdfundinsider}~\cite{io2017eosico}, EOSIO has become one of the largest Initial Coin Offering (ICO) projects (over \$4 billion). A recent report indicates that the average transaction volume of EOSIO within 24 hours has reached 57 million (80 million at peak)~\cite{io2017eosvolume}. By comparison, Ethereum has an average volume of only 717,000 transactions (1.3 million at peak) within 24 hours. 

ICO has become a new approach for many startups to raise funds. Different from traditional angel finance or venture capital, an ICO issuer raises cryptocurrencies by selling blockchain-based digital assets to users. In this way, cryptocurrencies can be interchanged with fiat money, consequently boosting the cryptocurrency economy. During this process, digital assets, also called \emph{tokens}, act as the programmable assets or access rights of participants in the blockchain. Tokens are essentially managed by smart contracts and underlying blockchains.  Owing to the high liquidity brought by the high transaction throughput and low confirmation latency, EOSIO tokens have become one of the most ideal choices for ICOs. Meanwhile, the waiver of trading fees in EOSIO is another attractive feature to stakeholders (e.g., token issuers and holders). 
%Both \emph{security} and \emph{liquidity} of the tokens are the most critical attributes to investors. Security can reduce the risk of investments while liquidity can provide investors with an arbitrage environment (e.g., highly frequent trading).

\begin{figure*}[t]
    \centering
	\includegraphics[width=0.8\textwidth]{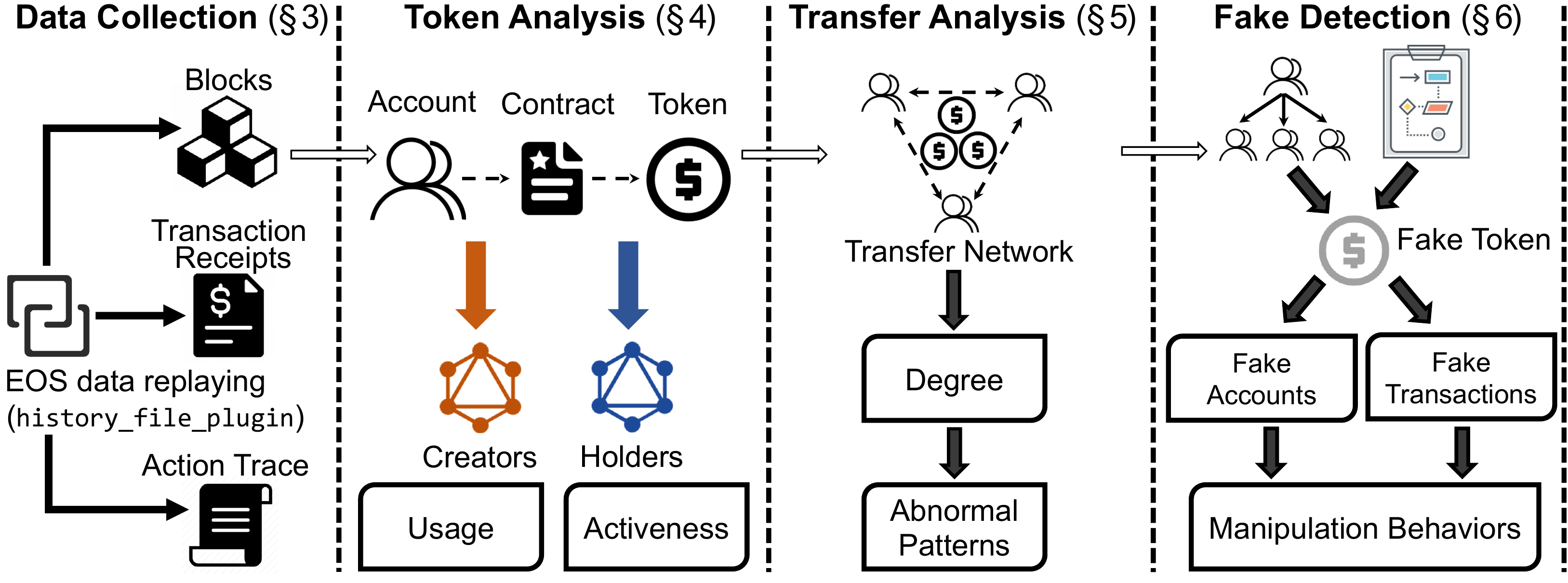}
	\vspace*{-0.25cm}
    \caption{An overview of the proposed framework to analyze EOSIO token data}
    \label{method}
\end{figure*}

\subsection{Motivation} 

Surprisingly, there are few studies on cryptocurrencies of EOSIO, considering its huge token transaction volume (i.e., more than 26.3 billion). An in-depth investigation of the EOSIO token ecosystem can help to reveal its internal mechanism and understand economic activities in EOSIO so as to demystify the token ecosystem. Despite a myriad of studies on EOSIO smart contracts~\cite{huang2020eosfuzzer,he2020security}, EOSIO vulnerabilities~\cite{huang2020characterizing}, and the Ethereum tokens~\cite{fenu2018ico,chen2020traveling,chen2019tokenscope,somin2018network,frowis2019detecting,victor2019measuring} (a more comprehensive literature survey to be given in Section~\ref{sec:RW}), there is \emph{no work to comprehensively investigate the EOSIO token ecosystem} to the best of our knowledge. 
%, especially for ERC20 tokens

To fill this gap, we conduct a systematic study on the EOSIO token ecosystem by performing extensive graph analysis on the entire on-chain EOSIO data. As shown in Fig.~\ref{method}, our study consists of four phases: (1) we collect the data of EOSIO and parse the token-related datasets; (2) we investigate the token ecosystem by constructing token creator graphs (TCGs), token contract creator graphs (TCCGs), and token holder graphs (THGs); (3) we analyze abnormal trading patterns by constructing token transfer graphs; (4) we propose an algorithm to detect suspicious tokens generated by fake users or fake transactions and analyze their corresponding manipulation behaviors.

\subsection{Contributions} 

In summary, we make the following contributions.
\begin{enumerate}
    \item To the best of our knowledge, we are the first to conduct a holistic measurement study on the whole EOSIO token ecosystem via graph analysis. After synchronizing the entire EOSIO data and gathering a large-scale dataset of all token-related transactions, we construct multiple graphs to characterize token creators, token contract creators, and token holders. The graph analysis offers an in-depth exploration of the entire EOSIO token ecosystem. We also compare EOSIO with Ethereum in token ecosystems. 
 
    \item After conducting the exploratory graph analysis, we analyze the tokens-transfer flows and observe some anomalous behaviors done by the accounts having large indegree or outdegree. These findings help us to identify abnormal trading patterns in EOSIO.
    
    \item We propose a fake-token detection algorithm to detect ``\textit{fake}'' tokens and identify manipulation behaviors. We extract several abnormal tokens and reveal their abnormal behaviors. Evaluation results further demonstrate the effectiveness of the algorithm.
\end{enumerate}

 %Through two dimensions of influence factors, we
The rest of the paper is organized as follows. After reviewing EOSIO and its internal mechanism in Section~\ref{Background}, we detail our study design and data collection in Section~\ref{sec:DC}. Section~\ref{sec:MC} then provides an overview of the EOSIO token ecosystem based on graph analysis. Section~\ref{sec:STA} next investigates the token transfer flows and identify some abnormal trading patterns. Section~\ref{sec:DTM} depicts the fake-token detection algorithm to identify the ``\textit{fake}'' tokens. After reviewing related work in Section~\ref{sec:RW}, we conclude the paper and outline future directions in Section~\ref{Conclusion}. 

\section{EOSIO in a Nutshell}\label{Background}
\subsection{Blockchain and EOSIO}
In general, a typical blockchain \cite{hamida2017blockchain,risius2017blockchain,zheng2018blockchain} is a globally-shared and distributed database, which is composed of a series of blocks containing transactions. A transaction refers to the interactive operation between users. And a block is constructed by transactions. Each block is confirmed by the entire network through a consensus protocol, such as PoW, PoS, and DPoS~\cite{cao2020performance,lepore2020survey,bach2018comparative,mingxiao2017review}. Participants in a blockchain system can read and write transactions in the blockchain database. There is no central authority in the blockchain. All the transactions are determined by the consensus protocol in a decentralized manner. As the core of blockchain technologies, the consensus protocol plays an important role in the development of the blockchain ecosystem.

As two main blockchain platforms, both Bitcoin and Ethereum are limited by PoW consensus protocols~\cite{meneghetti2020survey,NASIR2022136}. For example,  Bitcoin only supports 7 transactions per second while Ethereum supports 15 transactions per second. Different from Bitcoin and Ethereum, EOSIO adopts a more efficient consensus - DPOS - to scale the throughput to millions of transactions per second. Owing to the high scalability, EOSIO has gained huge popularity among users and developers. Another attraction of EOSIO to investors is the waiver of trading fees for any transactions, thereby greatly reducing the expenditure of high-frequency trading (such as arbitrage) for investors. 

The working flow of EOSIO is summarized as follows. 1) A \emph{user} first registers an EOSIO account, which can uniquely determine its identity. 2) The user interacts with the EOSIO blockchain through the invocation of smart contracts. The interaction is called an \emph{action} in EOSIO~\cite{xu2018eos}. 3) An EOSIO \emph{smart contract} written in C++ consists of contractual clauses, which can be invoked to be executed in EOSIO virtual machine (EOSVM)~\cite{io2017eosvm}, consequently generating a number of transactions to be stored in the EOSIO blockchain. 4) An EOSIO \emph{transaction} contains specific information about one or multiple users' actions, e.g., transferring tokens from one user to another. 

\subsection{Transaction, Action, and Account}
An EOSIO transaction consists of several actions, each of which represents an \emph{atomic} operation~\cite{xu2018eos}. Like traditional distributed database systems, the atomicity of a transaction means an indivisible set of actions in one transaction, i.e., either all of them are successful or none of them are successful. For example, a user namely Alice initiates an action consisting of (a) creating a new token named ``\texttt{\small TEST}'' and (b) transferring 10.0000 EOS\footnote{EOS is the token of EOSIO, similar to ether in Ethereum and BTC in Bitcoin.\label{eos-description}} to Bob. Both \textit{actions} (a) and (b) should occur either at the same time or none of them occurs. Both of the two actions are packaged into one transaction to be submitted to the EOSIO blockchain. As long as one of the actions fails, the entire transaction fails.

In EOSIO, a transaction is submitted by an account represented by a string with the length up to 12 characters. Creating a new account in EOSIO requires an existing account to pay a certain amount of EOS for RAM resources to store the account information. The existing account can be considered as the creator of the new account. Different from EOSIO, a new account (address) creation in Ethereum does not require the help of other accounts. This account-creation mechanism implies stronger relationships of EOSIO accounts than Ethereum. Therefore, it is worth investigating the relationships between EOSIO accounts while previous studies on Ethereum often ignore the relationship analysis. In Section~\ref{sec:DTM}, we propose an algorithm to detect ``\textit{fake}'' tokens and analyze the relationships of EOSIO accounts. 

\subsection{Smart Contract and Token}
Nowadays, most blockchain systems support smart contracts that run in virtual machines. Like other blockchain systems such as Ethereum, EOSIO smart contracts are also executed on top of EOSVM. In EOSIO, a smart contract written in C++ is first compiled to WebAssembly machine code (aka bytecode), which is then executed in EOSVM. Unlike Ethereum equipped with a \emph{gas} mechanism, EOSIO adopts a different resource-management mechanism, which limits the \emph{RAM}, \emph{CPU} and \emph{Bandwidth} resources for transaction execution to solve the \textit{halting problem}~\cite{burkholder1987halting,xu2018eos}. In EOSIO, an account can act as both a common user and a contract at the same time. When created, an account first acts as a common user. Authorized by its private key, it can interact with the blockchain on behalf of the user, such as sending tokens to other accounts. When this account is used to deploy a contract, the bytecode is stored into the account, which also serves as a contract. When a user invokes the contract, he/she initiates actions to the account. Consequently, the corresponding bytecode is executed in EOSVM to change the states of the blockchain. It is worth noting that the bytecode of an account can be updated (as long as owning its private key) in EOSIO, which is nevertheless not allowed in Ethereum.

In EOSIO, developers can easily use smart contracts to build their projects or DApps. Due to the waiver of trading fees and the simple development process of EOSIO DApps, many startups and ICOs raise funds by creating and issuing new tokens on the EOSIO platform. Any user can buy certain tokens of ICO DApps with EOS\textsuperscript{\ref{eos-description}}, which is the native token of EOSIO. A token that acts like a digital currency becomes a profitable asset for those shareholders of DApps. When the EOSIO \textit{mainnet} went live, a standard token protocol was introduced. As a result, the EOSIO token ecosystem has prospered rapidly and has soon become one of the largest token-selling platforms. Required by the EOSIO token standard, a token contract should consist of three functions: \texttt{\small create}, \texttt{\small issue}, and \texttt{\small transfer}. Required by the EOSIO token standard, a token contract should consist of three functions: \texttt{\small create}, \texttt{\small issue}, and \texttt{\small transfer}. Using this condition, we can filter all standard token contracts on the EOSIO \textit{mainnet}. If we parse the token-related transactions, we then can know how the tokens are transferred, where they go, and by whom they are held. It is worth mentioning that a token contract in EOSIO can create multiple tokens with different symbols and different contracts can create tokens with the same symbol. By contrast, this feature is also not allowed in Ethereum. Therefore, we uniquely mark a token with ``\textit{contract@symbol}'' in EOSIO.

\section{Study Design \& Data Collection}\label{sec:DC}
\subsection{Research Questions \& Study Methods}
In this paper, we aim to answer the following three research questions (RQs) when investigating the EOSIO token ecosystem:
{
\renewcommand{\theenumi}{RQ\arabic{enumi}}
\begin{enumerate}[wide, labelwidth=!, labelindent=0pt,noitemsep,topsep=0pt]
\item {\bf What are the market characteristics of the EOSIO token ecosystem?} % 图分析
  The EOSIO token ecosystem has huge market values due to its popularity and massive transactions. However, as far as we know, there is no study on investigating market characteristics by exploratory analysis on the EOSIO token data. To this end, we conduct a comprehensive graph analysis on tokens, holders, creators by constructing token creator graphs (TCGs), token holder graphs (THGs), and token contract creator graphs (TCCGs), respectively, accompanied by the relationship analysis. %We then identify the relationships between them.
\item {\bf Are there anomalous trading activities in the EOSIO token ecosystem?} % 特殊token分析，识别了一些异常的交易模式
  Tokens transferred in EOSIO reveal the trading flows, which can be used to identify trading activities, especially for those anomalous trading activities that may be a detriment to the EOSIO ecosystem. After analyzing token transfer graphs (TTGs) and characterizing the features, we find that some ``\textit{center}'' accounts have many \texttt{\small transfer} actions. We then analyze mutual trading activities and detect abnormal trading patterns.
\item {\bf Can we identify the tokens with fake users and transactions?} % 一个算法
  Although millions of token-related transactions occur in EOSIO, fake users or transactions commonly appear in EOSIO. Due to the waiver of trading fees of EOSIO, many token issuers intentionally increase both trading and user volumes of tokens with nearly no extra cost, thereby boosting the token popularity and gaining extravagant profits. To address this problem, we design an algorithm to detect these ``\textit{fake}'' tokens. We find that some identified cases can effectively reveal manipulation behaviors of tokens. 
\end{enumerate}
}

\subsection{Data Collection}
The collection of all the token-related actions requires replaying all transactions and gathering a large-scale dataset of all actions. However, the large transaction volume of EOSIO poses challenges in replaying transactions and efficiently obtaining the entire on-chain data. Although the EOSIO development team offers the client \texttt{\small Nodeos} and several plugins, like \texttt{\small state\_history\_plugin} and \texttt{\small mongo\_db\_plugin}, the official plugins severely slow down the replay procedure due to parsing and insertion operations of raw data to databases. These plugins collect the raw data when replaying transactions, and then parse them into the well-formatted data for some database engines (i.e., PostgreSQL and MongoDB). Finally, the formatted data are inserted into the database according to certain indexes (with the purpose of fast query). Data insertion operations take extra time during the replaying procedure. Meanwhile, data parsing and insertion operations are conducted serially and may affect each other, thereby further slowing down the replaying procedure. 

To address these challenges, we develop a new data-replaying plugin - \texttt{\small history\_file\_plugin} to collect raw data and write them into \textit{Memory Buffer} during the replaying procedure. Then, another thread asynchronously reads the data from \textit{Memory Buffer}, serializes, and finally saves them directly as JSON files. Since the subsequent data preprocessing is conducted on these files without affecting the replaying procedure, \texttt{\small history\_file\_plugin} allows data collection and data processing to be carried out simultaneously, consequently speeding up data collection. Our plugin greatly saves time in collecting the entire on-chain data in contrast to official plugins of EOSIO. For example, our plugin takes only 1/7 time to synchronize the first 20 million blocks, compared with the official plugins of EOSIO\footnote{Our plugin is expected to obtain even better results than the official plugins for the entire EOSIO dataset because of no insertion operations to databases.}. 

\begin{table}[t]
  \centering
  \setlength{\belowcaptionskip}{0.1cm}
  \caption{EOSIO Token Data: Block \#1 to \#134,999,999}
  \label{table:token-data}
 \scriptsize
  \begin{tabular}{c|c|cl}
    \toprule
    \textbf{Category} & \textbf{Approximate size of Dataset} &\textbf{Row Count}\\
    \midrule
     \texttt{token create actions} & \texttt{944 KB}  & \texttt{5,598}\\
     \texttt{token issue actions} &  \texttt{40.42 GB} & \texttt{253,711,757}\\
     \rowcolor{LightCyan}
     \texttt{token transfer actions} & \texttt{4.23 TB} & \texttt{26,311,585,008}\\
     \texttt{account creation actions} &  \texttt{244.62 MB} & \texttt{1,332,669}\\
\bottomrule
  \end{tabular}
\end{table}

\textbf{EOSIO Token Data Summary:} We have launched \textit{Nodeos} and our own \textit{history\_file\_plugin} to run an EOSIO full node and replay all the transactions (up to 134,999,999 blocks) to get the entire on-chain data (including blocks, transaction receipts, action traces) from June 8, 2018 to Aug. 5, 2020. According to the token standard defined by EOSIO, we filter out all standard tokens and extract the token-related actions covering creation, issuance, and transfer. Table~\ref{table:token-data} summarizes the EOSIO token data, which obviously has much larger volumes than Ethereum~\cite{10.1145/3442381.3449916}. More details about the dataset are shown below.

\textbf{Token Information:} In EOSIO, a contract that contains three standard functions of \texttt{\small create}, \texttt{\small issue}, and \texttt{\small transfer} can be regarded as a standard token contract. According to this feature, we filter out 2,047 contracts to be considered as standard token contracts, which have created and issued 5,598 tokens. For these 5,598 tokens, we collect the data including \texttt{\small create} actions, \texttt{\small issue} actions, and \texttt{\small transfer} actions for each token. A \texttt{\small create} action represents that a user creates a token through a token contract. An \texttt{\small issue} action represents that an issuer issues some tokens to a user directly (also known as Token \texttt{\small Airdrop}), while a \texttt{\small transfer} action represents that a user sends some tokens to another user. There are 253,711,757 token \texttt{\small issue} actions that were submitted by 2,140 issuers and 26,311,585,008 token \texttt{\small transfer} actions that occurred in 1,332,669 holding accounts. Table~\ref{table:token-t-format} illustrates an example of a \texttt{\small transfer} action for helping readers further understand the format of the dataset. And the format of the other two actions is simple and similar to that of the \texttt{\small transfer} action.

\textbf{Account Information:} In EOSIO, creating a new account requires an existing creator to pay a certain amount of EOS. There are 2,096,840 distinct accounts, which were created by only 48,691 account creators. We filter out all account creation actions and extract their relationship as the format shown in Table~\ref{table2}.
\begin{table*}[h]
  \centering
  \setlength{\belowcaptionskip}{0.1cm}
  \caption{Token Transfer Format}
  \label{table:token-t-format}
 \footnotesize
  \begin{tabular}{c|c|cl}
    \toprule
    \textbf{Category} & \textbf{Description} &\textbf{Data}\\
    \midrule
     \texttt{txid} & \texttt{transaction\ id}  & \texttt{07fc627668a471c3d...}\\
     \texttt{block\_time} &  \texttt{block\ timestamp} & \texttt{2018-06-10T14:23:39.000}\\
     \texttt{contract@symbol} & \texttt{the token contract and token symbol} & \texttt{eosnowbanker@EOSNOW}\\
     \texttt{from} &  \texttt{token\ sender} & \texttt{eosnowbanker}\\
     \texttt{to} & \texttt{token\ receiver}& \texttt{gqztamzsg4ge} \\
     \texttt{quantity} & \texttt{the\ amount\ of\ token}& \texttt{10000.0000\ EOSNOW} \\
     \texttt{memo} & \texttt{transfer\ memo}& \texttt{Now\ is\ now\ Now} \\
\bottomrule
  \end{tabular}
 % \vspace*{-0.5cm}
\end{table*}

\begin{table}[h]
  \centering
  \setlength{\belowcaptionskip}{0.1cm}
  \caption{Account Creation Format}
  \label{table2}
  \scriptsize
  \begin{tabular}{c|c|cl}
    \toprule
    \textbf{Category} & \textbf{Description} &\textbf{Data}\\
    \midrule
     \texttt{txid} & \texttt{transaction\ id} & \texttt{245786e9d77657a5e...}\\
     \texttt{block\_time} &  \texttt{block timestamp}& \texttt{2018-06-12T17:05:16.500}\\
     \texttt{creator}& \texttt{the\ name\ of\ the\ creator} & \texttt{hezdqmbygyge}\\
     \texttt{name} & \texttt{the\ new\ account\ name} & \texttt{iloveuzi3344}\\
\bottomrule
  \end{tabular}
%  \vspace*{-0.5cm}
\end{table}

\section{Token Analysis}\label{sec:MC}
In this section, we provide an overview of the EOSIO token ecosystem. We concentrate our study on the tokens and explore the characteristics of the token ecosystem. Based on the analysis, we obtain the following findings.
\begin{itemize}[wide, labelindent=0pt,labelindent=0pt,noitemsep,topsep=0pt]
    \item \textbf{Finding 1}: Despite a huge volume of token \texttt{\small transfer} actions (exceeding 26.3 billion), most of the tokens are ``\textit{silent}''. Specifically, nearly 80\% of the tokens are traded less than 100 times and only 1\% of the tokens cover more than 90\% of the total token volume. 
    \item \textbf{Finding 2}: Tokens can be created by an account through one or multiple contracts. Some accounts create a large number of tokens through one contract possibly because of testing or ``just for fun''. On the contrary, few creators create multiple tokens through multiple contracts.
    \item \textbf{Finding 3}: A small number of accounts (might be some exchanges) hold a large number of tokens while most accounts hold a small number of tokens. Similarly, most tokens are held by a small number of holders while only a small number of holders have a large number of tokens.
\end{itemize}

\subsection{Token Activeness and Token Usage}\label{MC_TA}
\begin{figure}[b]
\centering 
\subfigure[\scriptsize Distribution of token activeness]{
\centering
\includegraphics[height=1.2 in]{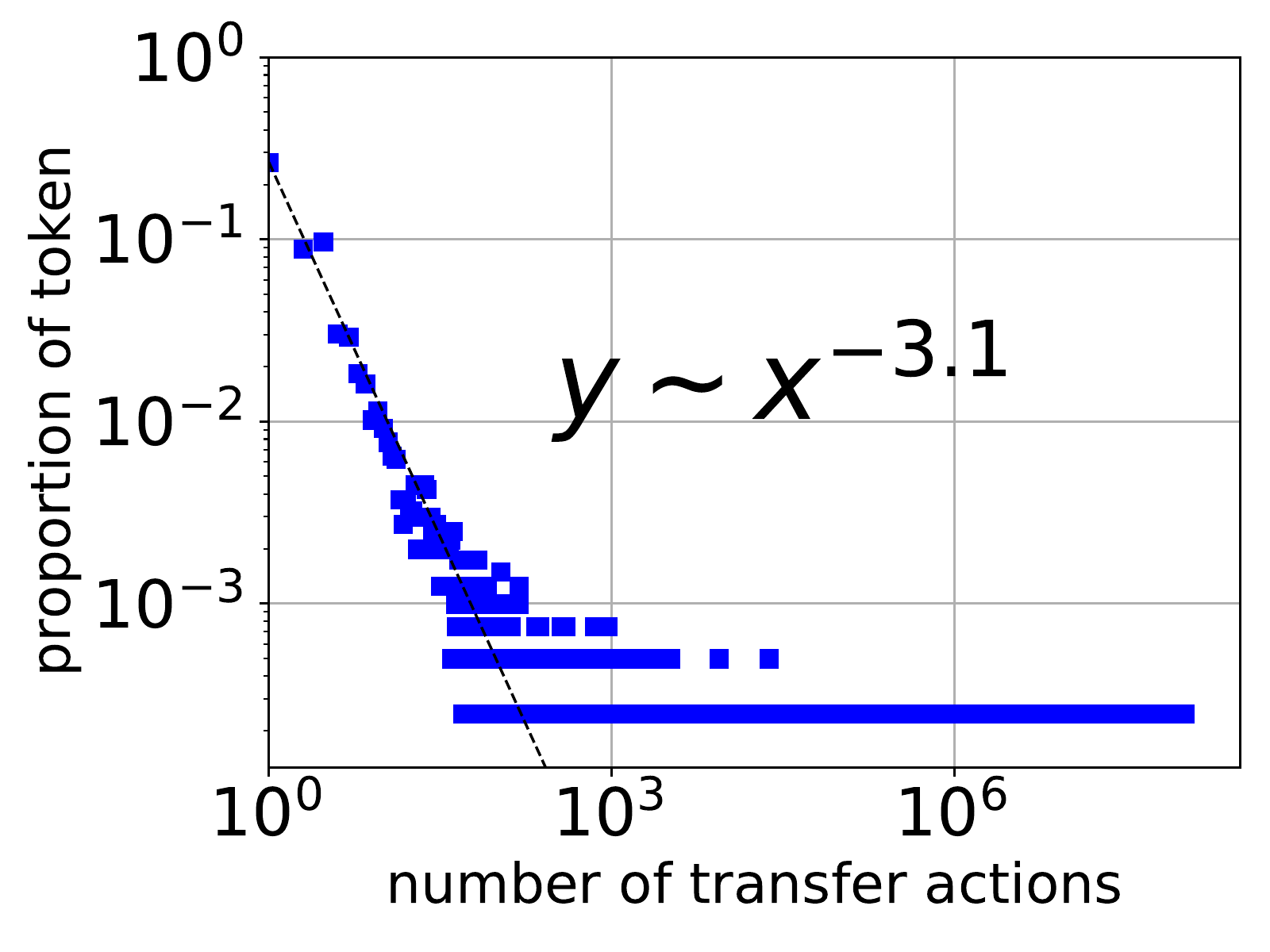}\label{tokenactiveness}
}
\hspace{3mm}
\subfigure[\scriptsize Word cloud of memos]{
\centering
\includegraphics[height=1.2 in]{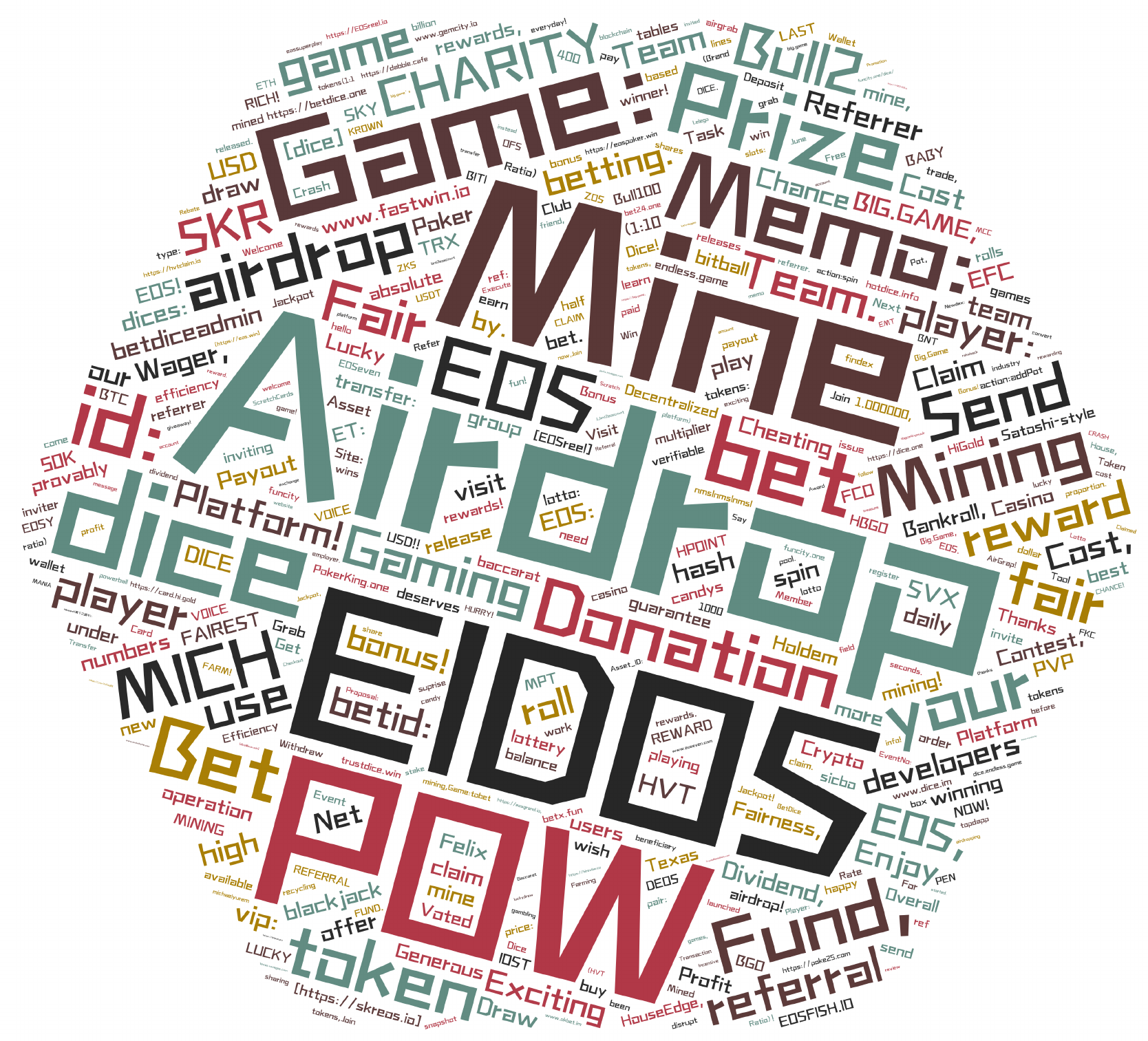}\label{tu-2}
}
\vspace*{-0.25cm}
\caption{Token activeness and usage}
\label{token_activeness_and_usage}
\end{figure}

As an important measure on the health of the token ecosystem, the degree of the \emph{token activeness} reveals the network status and the availability of the ecosystem. We first define the \emph{token activeness} of a token as the number of its \texttt{\small transfer} actions. This metric has been used as an important indicator for ranking in many DApp websites (e.g., DappReview). We then plot the distribution of the token activeness in Fig.~\ref{tokenactiveness}, from which the \emph{Matthew effect}~\cite{merton1968matthew} can be observed. Nearly 27.9\% of the tokens have never been transferred, and 78.9\% of the tokens are transferred less than 100 times. Meanwhile, 1\% of the tokens cover more than 90\% of the total volume, thereby further confirming the existed Matthew effect. This  result indicates that most tokens do not succeed from the perspective of users' activity. In other words, there are only a few \emph{active} tokens while most of them are \textit{silent}. To further analyze the token activeness, we plot the fitted line for the distribution through $y{\sim}x^{-\beta}$ as shown in Fig.~\ref{tokenactiveness}. The larger $\beta$ leads to the smaller degree of the token activeness.

\renewcommand\subfigcapskip{-0.5ex}

% \renewcommand\subfigcapskip{-0.5ex}
% \begin{figure*}[h]
% \begin{minipage}[h]{0.612\linewidth}
% \centering
% \includegraphics[width=1.4in]{figure/tokenActive.pdf}
% %\vspace*{-0.3cm}
% \caption{\textbf{Distribution of token activeness}}
% \label{tokenactiveness}
% \vspace*{-0.3cm}
% \end{minipage}
% \begin{minipage}[h]{0.382\linewidth}
% \centering
% \includegraphics[height=1.44 in]{figure/wc.jpg}
% \caption{\textbf{Word cloud statistics of memos}}
% \label{wordcloud}
% \vspace*{-0.3cm}
% \end{minipage}
% \end{figure*}

%The larger $\beta$ leads to the smaller degree of the token activeness.

% In order to reveal the characteristics of active tokens, we analyze the top-5 most active tokens according to the number of \texttt{\small transfer} actions, as shown in Table~\ref{table3}. 
% We obtain the identities of these popular tokens through searching either blogs or some DApp websites so as to help readers understand their usage. 

% \vspace*{-0.25cm}
\begin{table*}[h]
	\centering
	\setlength{\belowcaptionskip}{0.1cm}
	\caption{Top-5 Tokens according to Token Activeness}
	\label{table3}
	\footnotesize
	\renewcommand{\arraystretch}{0.9}
	\begin{tabular}{c|c|cl}
		\toprule
		\textbf{Tokens}             & \textbf{Number of \texttt{transfer} actions} & \textbf{Identities}              \\
		\midrule
		%  \hline   %  or \cline{col1-col2}
		\rowcolor{LightCyan}
		\texttt{eidosonecoin@EIDOS} & \texttt{23,484,345,961}                      & \texttt{Airdrop for DDoS attack} \\
		\texttt{eosiopowcoin@POW}   & \texttt{1,793,696,754}                       & \texttt{CPU Mining}              \\
		\texttt{betdicetoken@DICE}  & \texttt{103,865,132}                         & \texttt{BetDice, Gambling Game}  \\
		\texttt{bgbgbgbgbgbg@BG}    & \texttt{74,043,095}                          & \texttt{BigGame, Gambling Game}  \\
		\texttt{mine4charity@MICH}  & \texttt{67,464,460}                          & \texttt{CPU Mining}              \\
		\bottomrule
	\end{tabular}
\end{table*}

Table~\ref{table3} lists the top-5 most active tokens. We find that \texttt{\small EIDOS} is the most active token with up to 23 billion \texttt{\small transfer} actions. According to ``Blocking.net'', \texttt{\small EIDOS} leads to a token \texttt{\small airdrop} feast aimed at exposing the defects of EOSIO's resource management and even ``\textit{killing}'' EOSIO~\cite{eosisdead}. Anyone who transfers 0.0001 EOS to contract \texttt{\small eidosonecoin} can then receive 0.0001 EOS as well as some EIDOS tokens from \texttt{\small eidosonecoin}. To gain more EIDOS tokens, many users submitted a large number of \texttt{\small transfer} actions to \texttt{\small eidosonecoin}, thereby consuming substantial CPU resources. At the peak, the CPU resources consumed by \texttt{\small eidosonecoin} occupy 60\% of the entire network according to DAppTotal~\cite{DAppTotal}, consequently causing users to be unable to transfer money normally and leading to the dysfunction of other DApps. This abnormal behavior can be regarded as a DDoS attack on the EOSIO \texttt{\small mainnet}.  \texttt{\small POW} and \texttt{\small MICH} have a similar operating model (commonly known as CPU Mining) to \texttt{\small EIDOS}. All these projects caused some harm to EOSIO's resource management. Acting as tokens for gambling games, both \texttt{\small DICE} and \texttt{\small BG} have been operating gambling markets since September 2018. From the popularity of these two tokens, we speculate the popularity of \textit{gambling and gaming} in EOSIO owing to the waiver of trading fees of EOSIO in contrast to other blockchain platforms.

% Acting as tokens for gambling games, \texttt{\small DICE} (created on Sep. 20, 2018) and \texttt{\small BG} (created on Sep. 18, 2018) are all operating gambling markets. From the popularity of these two tokens, we speculate the popularity of \textit{gambling and gaming} in EOSIO owing to the waiver of trading fees of EOSIO in contrast to other blockchain platforms.

%The operating model (commonly known as CPU Mining) of

It is difficult to verify the functionality of each token since most tokens do not have any relevant information except for some well-known ones. Thus, we go through all the \texttt{\small transfer} actions of each token and collect the \textit{memo} of each action. These memos usually imply the purposes of the actions (e.g., betting) and the potential identities of the \emph{senders}. Fig.~\ref{tu-2} depicts the word cloud of the memos of EOSIO tokens. The most common word is ``\texttt{\small Airdrop}'', indicating that the token \texttt{\small airdrop} occurs the most frequently in EOSIO. Meanwhile, the words ``\texttt{\small EIDOS}'', ``\texttt{\small POW}'', ``\texttt{\small Mine}'' indicate the prevalence of CPU Mining. Other frequent words include ``\texttt{\small Bet}'', ``\texttt{\small Game}'', ``\texttt{\small Prize}'' (related to gambling and game actions), further confirming the huge popularity of both gambling apps and games in EOSIO.
% Table~\ref{table3} also lists the identities of these popular tokens so as to help readers understand the usage of these active tokens. We obtain them mainly through searching either blogs or some DApp websites.  We next analyze the relationships between these tokens, their creators, and their users.

% eidosonecoin+EIDOS
% 23484345961
% eosiopowcoin+POW
% 1793696754
% betdicetoken+DICE
% 103865132
% bgbgbgbgbgbg+BG
% 74043095
% mine4charity+MICH
% 67464460
% bitpietokens+EUSD
% 50018475
% ipsecontract+POST
% 38589244
% pokereotoken+PKE
% 38479654

\subsection{Token Creators}\label{MC_TC}
\begin{figure}[h]
\centering 
\subfigure[\scriptsize One contract creates one token]{
\centering
\includegraphics[height=0.82 in]{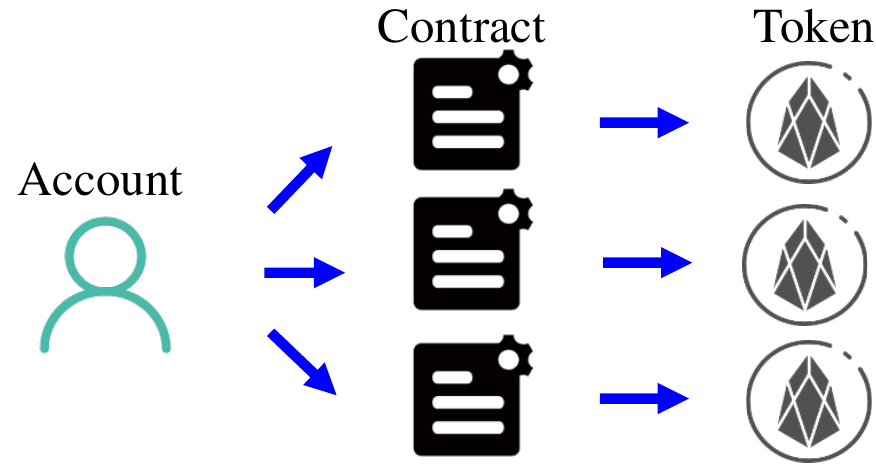}\label{ra-1}
}
\hspace{4.5mm}
\subfigure[\scriptsize One contracts creates multiple tokens]{
\centering
\includegraphics[height=0.82 in]{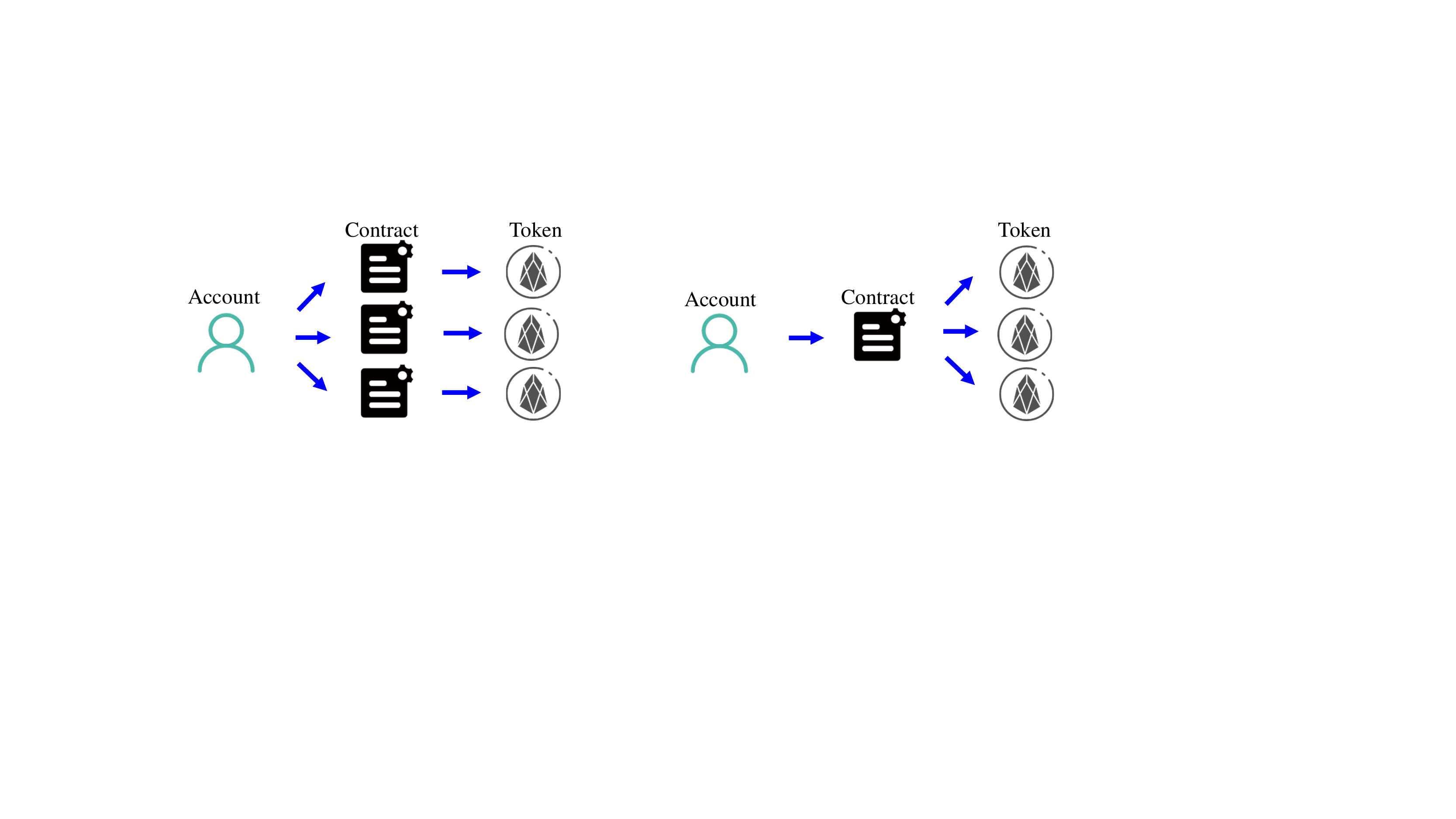}\label{ra-2}
}
\vspace*{-0.25cm}
\caption{Relationship between accounts and tokens}
\label{relationship_token_user}
\end{figure}

Different from Ethereum, in which one token contract can create only one token, a contract in EOSIO can create one or multiple tokens, as shown in Fig.~\ref{relationship_token_user}. In the first case, an account is able to deploy one token contract, which can be invoked to create multiple tokens, as shown in Fig.~\ref{ra-1}. Thus, a contract in EOSIO can be reused for token creation. In the second case, an account can create multiple tokens through multiple contracts, as shown in Fig.~\ref{ra-2}. EOSIO allows different contracts to create tokens with the same name (symbol) while Ethereum disallows this feature.

To investigate the relationships between tokens and accounts, we focus on the number of tokens created by each account. We introduce $\mathsf{TCG}$ to investigate token creators as follows:
{ 
$$\mathsf{TCG}=(V_\text{at},E_\text{at},D),E_\text{at}=\{(v_i,v_j,d)|v_i,v_j \in V_\text{at} ,d  \in D\},$$}%
where $V_\text{at}$ is a set of accounts and tokens, and $E_\text{at}$ is a set of edges. Each edge $(v_i,v_j,d)$ indicates the creation relationship between an account $v_i$ and a token $v_j$ with a timestamp $d$ (between June 10, 2018 and Aug. 5, 2020, the same below). To explore whether there are tokens with the same symbol, we use ``\textit{symbol}'' instead of ``\textit{contract@symbol}'' to mark a token node in $\mathsf{TCG}$.

\begin{figure}[h]
% \vspace{-0.5cm}
\centering 
\subfigure[\scriptsize $\mathsf{TCG}$]{
\centering
\includegraphics[height=1.2 in]{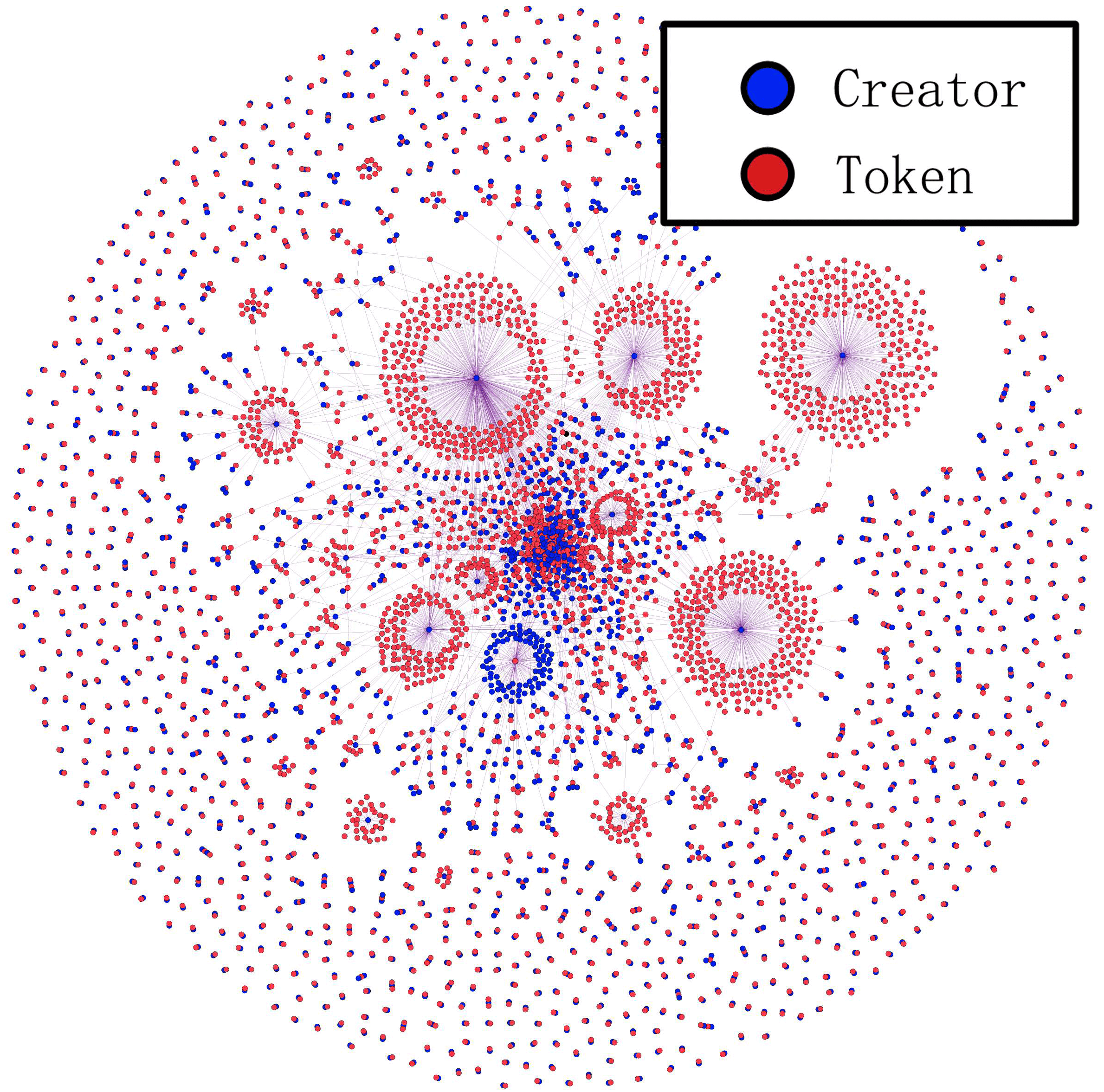}\label{TCG}
}
\quad
\subfigure[\scriptsize Outdegree distribution of $\mathsf{TCG}$]{
\centering
\includegraphics[height=1.2 in]{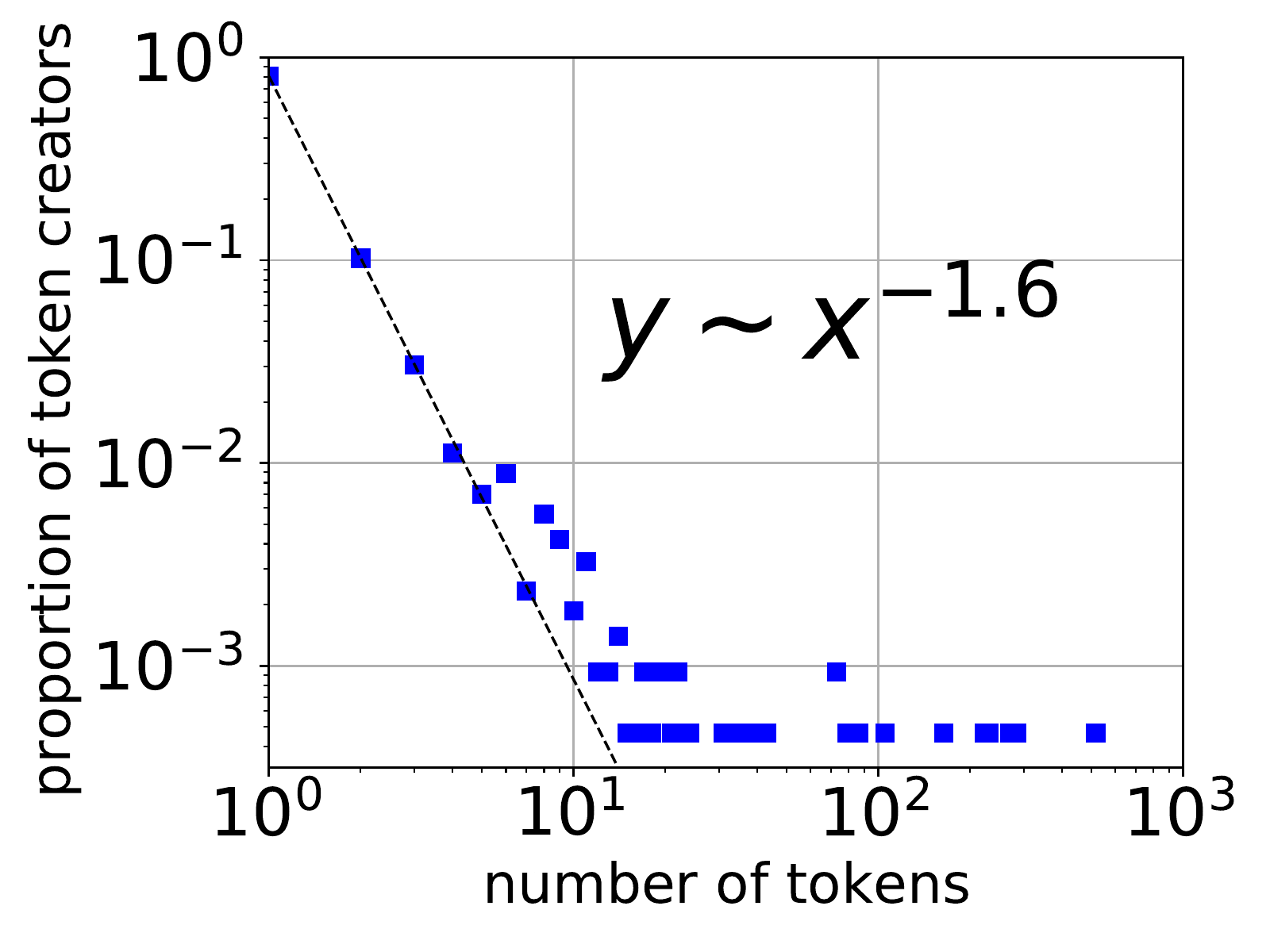}\label{outdegree-TCG}
}
\vspace*{-0.25cm}
\caption{Visualization of Token Creators ($\mathsf{TCG}$)}
\end{figure}

Fig.~\ref{TCG} illustrates the $\mathsf{TCG}$ constructed from our collected dataset, where creators are marked in blue and tokens are marked in red. We observe from Fig.~\ref{TCG} that a small number of accounts create a large number of tokens (i.e., one blue node is circled by many red nodes) while most of the accounts only create one or two tokens. Meanwhile, we also find an abnormal phenomenon, in which one red node is circled by many blue nodes. It can be explained by the fact that the tokens with the same symbol are created by multiple creators. For example, we find that there are 158 tokens named \texttt{\small EOS} being created by 158 accounts through different contracts. One reason why creators prefer the symbol \texttt{\small EOS} may lie in \texttt{\small EOS} being the native token of EOSIO so as to attract more attention. Moreover, some attackers also create the token named \texttt{\small EOS} to initiate the ``\textit{fake EOS}'' attacks to some vulnerable contracts and steal tokens~\cite{huang2020characterizing}.

To further analyze the characteristics of the $\mathsf{TCG}$, we plot the \emph{outdegree distribution} of creators in Fig.~\ref{outdegree-TCG}. The \emph{outdegree distribution} essentially indicates the number of tokens created by the creators. Fig.~\ref{outdegree-TCG} reveals a strong power-law distribution reflecting a small number of nodes with a large outdegree. Moreover, nodes with smaller outdegree in the token ecosystem account for the majority. For example, nearly 80.6\% of the creators only created one token and 95.7\% of the creators created no more than 5 tokens. In addition, the account who created the most number of tokens monopolized 517 tokens, leading to a severe polarization of distribution.

Besides the relationship between tokens and creators (as analyzed in $\mathsf{TCG}$), we next analyze the relationship between tokens and token contracts. We define $\mathsf{TCCG}$ as follows:
{ 
$$\mathsf{TCCG}=(V_\text{tc},E_\text{tc},D),E_\text{tc}=\{(v_i,v_j,d)|v_i,v_j \in V ,d  \in D\},$$}%
where $V_\text{tc}$ is a set of the token contracts and tokens and $E_\text{tc}$ is a set of edges. An edge $(v_i,v_j,d)$ represents that a token $v_i$ is created by a token contract $v_j$ on timestamp $d$. $\mathsf{TCCG}$ has a similar distribution to $\mathsf{TCG}$, implying that both $\mathsf{TCG}$ and $\mathsf{TCCG}$ have homologous relationships. A token often has the same account for its creator and its contract (as mentioned in Section~\ref{Background}, an account can act as both a user and a contract). Meanwhile, we also find that creators prefer using the same contract rather than using multiple contracts to create multiple tokens. The reusability of token contracts brings convenience and saves costs since creators do not need to deploy another contract.

\begin{figure}[h]
    \centering
    \includegraphics[width=0.48\textwidth]{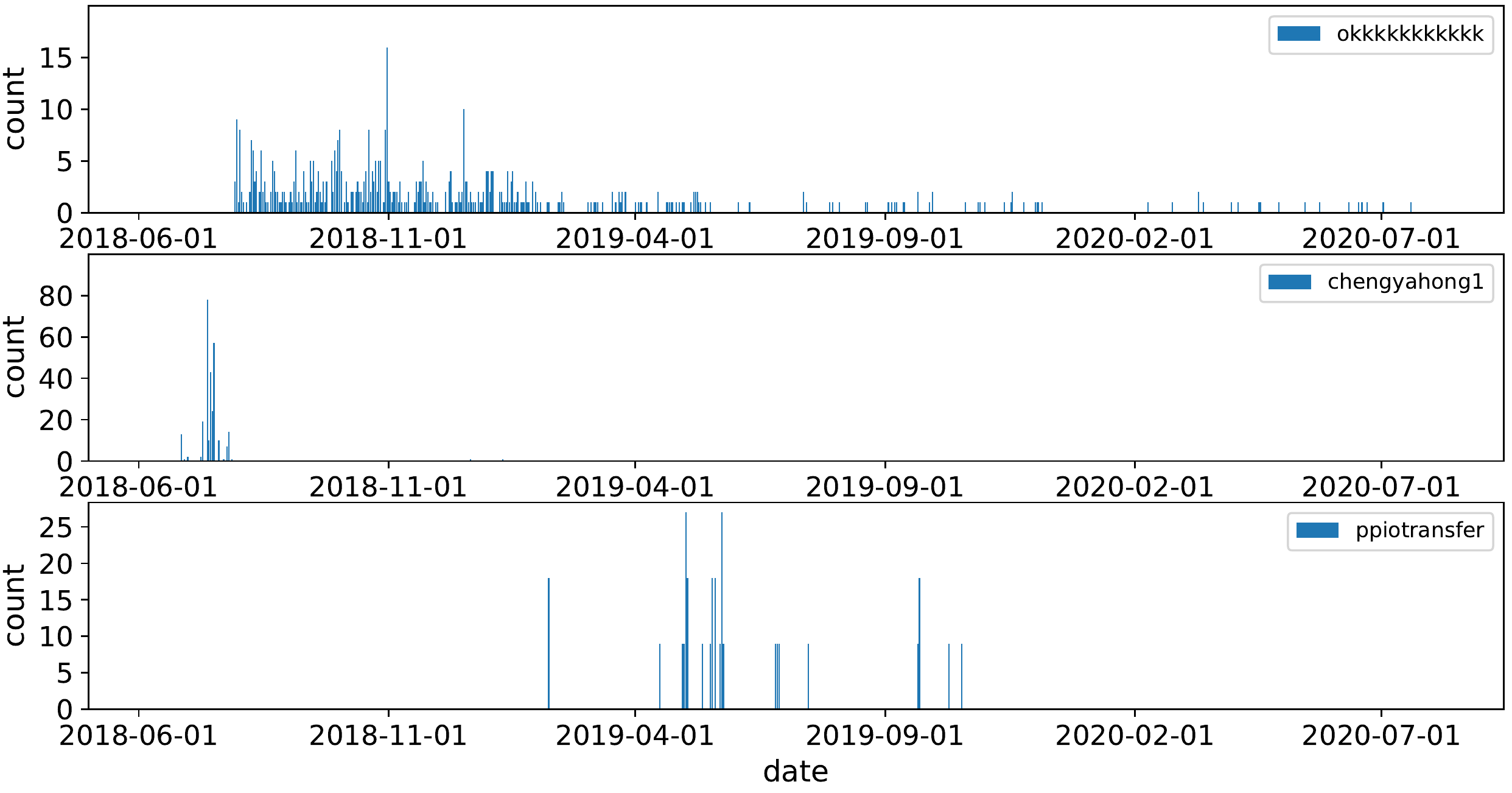}
    \vspace*{-0.25cm}
    \caption{Distribution of token creation over time}
    \label{TCG-Top-3}
\end{figure}

\textbf{Who Created The Most Tokens?}
%\label{MC_WCTMT}
We then concentrate on the accounts that created the most tokens and summarize the relevant characteristics of top-3 creators. Account \texttt{\small okkkkkkkkkkk} is the creator with the most number of tokens (517 tokens). By carefully analyzing all actions related to account \texttt{\small okkkkkkkkkkk}, we find that \texttt{\small okkkkkkkkkkk} usually receives \texttt{\small eosbtextoken@BT} tokens from many accounts and then sends different tokens (e.g., \texttt{\small USDS}, \texttt{\small DNA}, \texttt{\small LOVEYOU}) to these accounts. It implies that \texttt{\small okkkkkkkkkkk} is probably an intermediary between \texttt{\small BT} token and other tokens, thereby providing a decentralized service for token exchange. The second-rank creator \texttt{\small chengyahong1} creates 284 tokens while the third-rank creator \texttt{\small ppiotransfer} creates 270 tokens. Our further analysis shows that both these two creators often issue or send the tokens created by themselves to the same account, implying that they may create tokens for testing or just for fun. To reveal the differences between these creators, we study the distribution of token creation over time after counting the number of tokens created by these three creators everyday. As shown in Fig.~\ref{TCG-Top-3}, \texttt{\small okkkkkkkkkkk} has been continuously creating tokens. Account \texttt{\small okkkkkkkkkkk} has tracked the new initiated projects as well as their tokens and created new tokens and token pairs to meet the needs for the token exchange of ICOs. This further confirms the identity of \texttt{\small okkkkkkkkkkk}, who is a token intermediary. On the contrary, both accounts \texttt{\small chengyahong1} and \texttt{\small ppiotransfer} only sporadically create tokens.

\subsection{Token Holders}\label{MC_TH}
We further investigate the holders of the tokens and identify their characteristics. To this end, we define and construct $\mathsf{THG}$ as follows:
{$$\mathsf{THG}=(V_\text{th},E_\text{th},w),E=\{(v_i,v_j,w)|v_i,v_j \in V, w  \in (0,1]\},$$}%
where $V_\text{th}$ is a set of tokens and holders, and $E_\text{th}$ is a set of edges, in which each edge indicates the holding relationship between a holder $v_i$ and a token $v_j$. Note that each edge is also associated with a weight $w$, indicating that $v_i$ holds $w$ shares of token $v_j$.

\begin{figure}[h]
\centering 
\subfigure[\scriptsize $\mathsf{THG}$]{
\centering
\includegraphics[height=0.90 in]{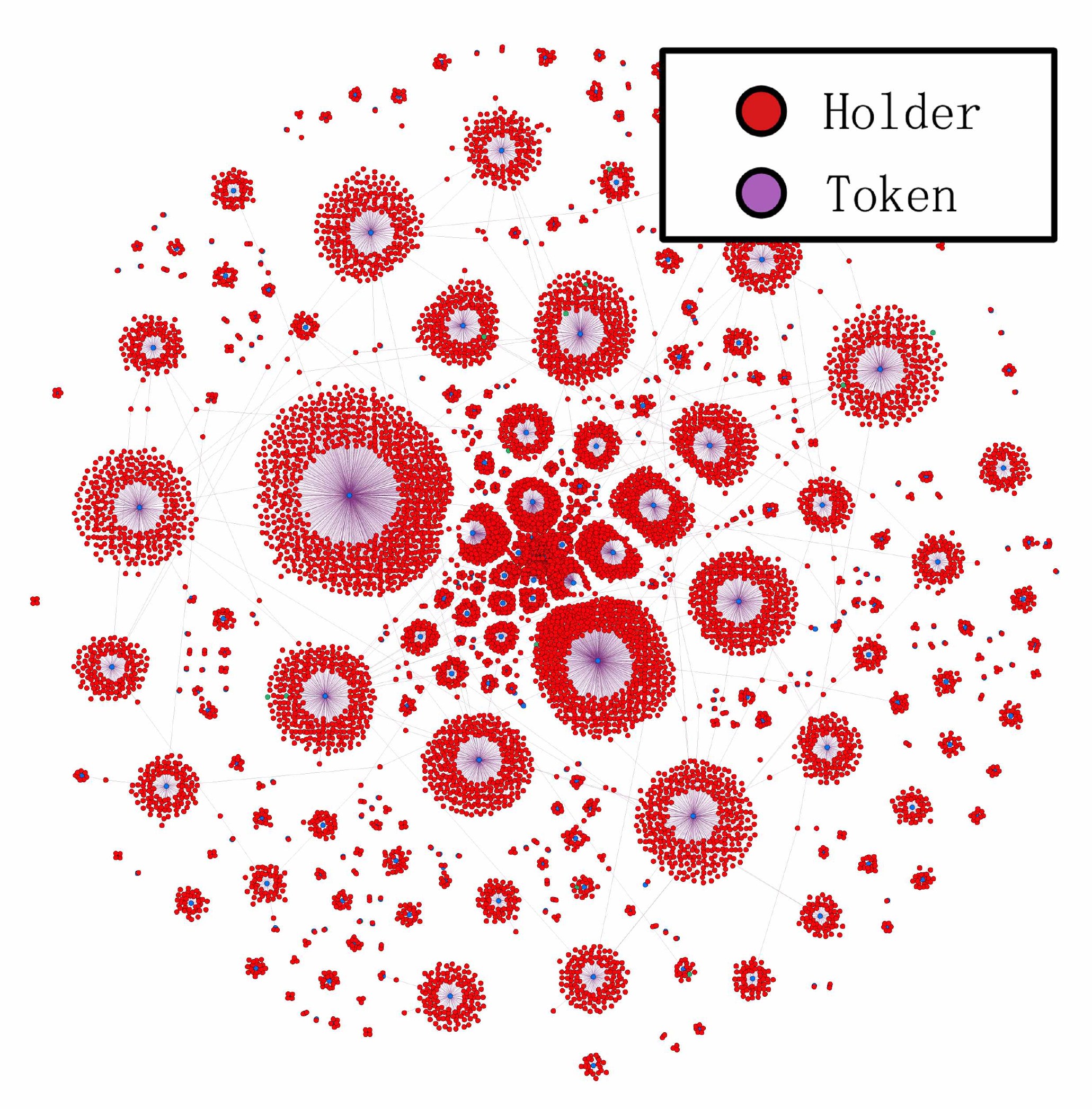}\label{THG}
}
\subfigure[\scriptsize $\mathsf{THG}$ indegree distribution]{
\centering
\includegraphics[height=0.85 in]{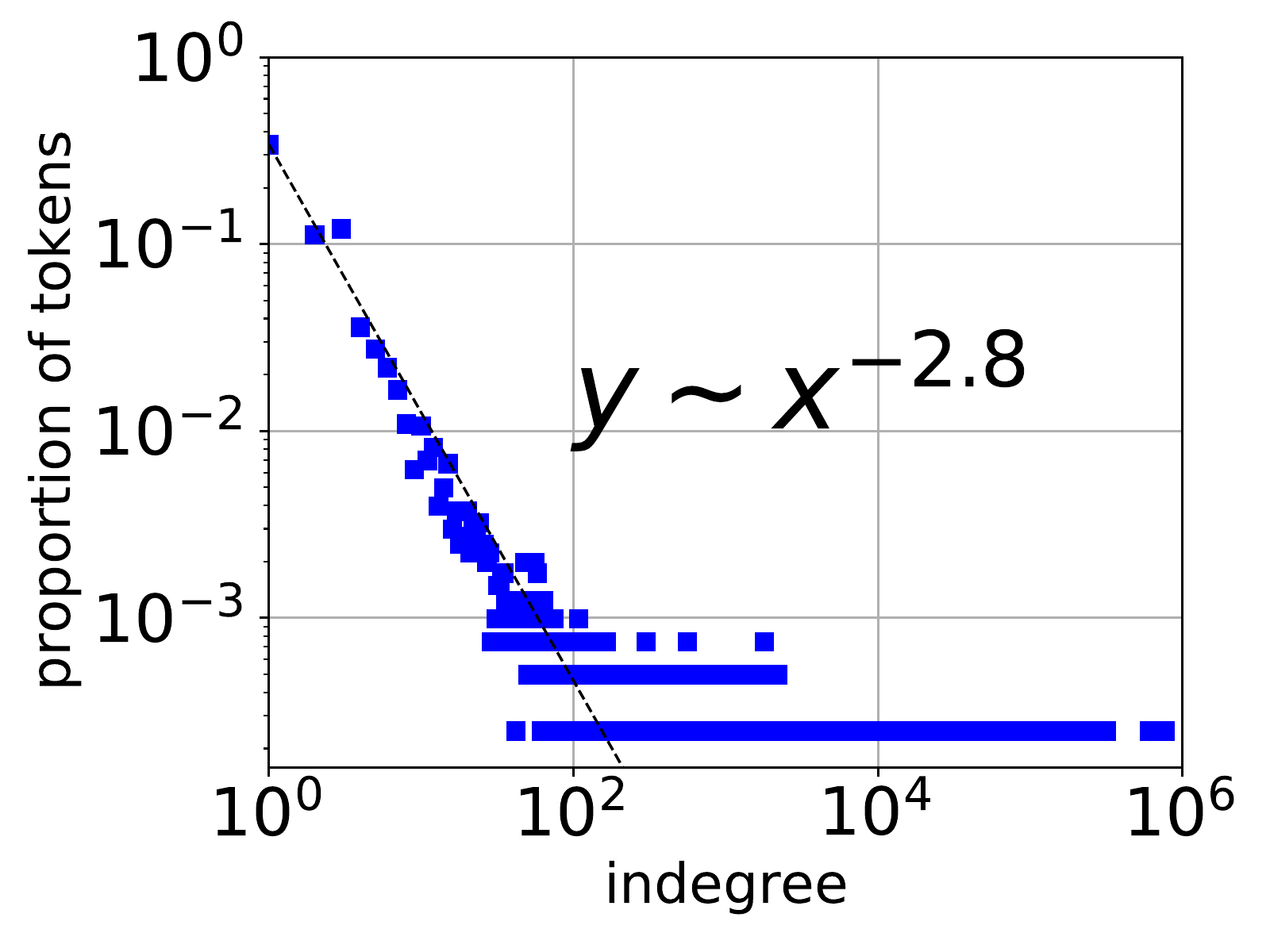}\label{indegree_THG}
}
\subfigure[\scriptsize $\mathsf{THG}$ outdegree distribution]{
\centering
\includegraphics[height=0.85 in]{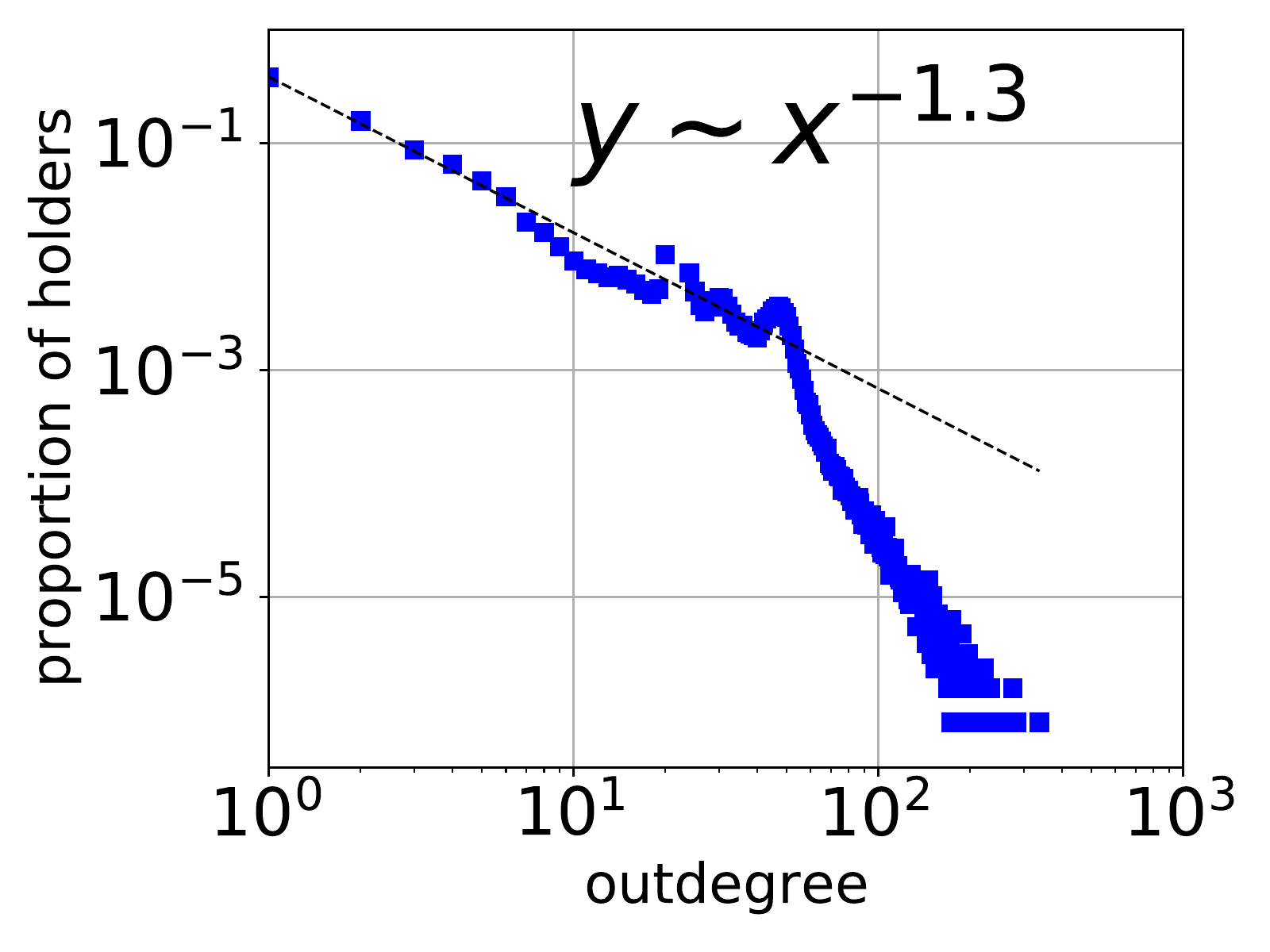}\label{outdegree_THG}
}
\vspace*{-0.25cm}
\caption{Visualization of Token Holders}
\label{fig:THG}
\end{figure}

Fig.~\ref{fig:THG} presents an exploratory analysis of $\mathsf{THG}$. Fig.~\ref{THG} first gives the visualization of $\mathsf{THG}$, in which the purple nodes denote the tokens and the red nodes denote the holders. Fig.~\ref{THG} reveals that several popular tokens are owned by many holders while most of the tokens are still possessed by very few holders. 
Figs.~\ref{indegree_THG} and~\ref{outdegree_THG} show the \emph{indegree} and \emph{outdegree distribution} of $\mathsf{THG}$, respectively. The indegree of a token in $\mathsf{THG}$ means the number of its holders while the outdegree of a holder is the number of tokens that he/she holds. We observe an approximate power-law distribution, i.e., there are lots of small-degree nodes while few large-degree nodes.

\renewcommand\subfigcapskip{-0.5ex}

{\textbf{Who Hold the Most Tokens?}}
%\label{MC_WHTMT}
Analyzing the \emph{outdegree distribution} of the holders, we find that there are 1,332,669 holders and 35.63\% of them hold only one token in the token ecosystem. Moreover, 84.88\% of the holders possess fewer than 20 tokens. Table~\ref{table5} lists the top-3 holders possessing the most tokens, and \textbf{\small Invocation} in Table~\ref{table5} represents the number of \texttt{\small transfer} actions involving a holder, who is either the \emph{sender} or the \emph{receiver}.

\begin{table}[t]
  \centering
  \setlength{\belowcaptionskip}{0.1cm}
  \caption{Top-3 Accounts of $\mathsf{THG}$ Using Degree Centrality}
  \label{table5}
  \scriptsize
\renewcommand{\arraystretch}{0.9}
  \begin{tabular}{c|c|c|cl}
    \toprule
    \textbf{Accounts} & \textbf{Outdegree} &  \textbf{Invocation} &  \textbf{Identities}\\
    \midrule
     \rowcolor{LightCyan}\texttt{newdexiofees} & \texttt{338} & \texttt{3,873,999}&  \texttt{decentralized exchange} \\
     \texttt{5lisqkvt1n2q} & \texttt{284}  & \texttt{3,430} &  \texttt{token speculator, arbitrageur}\\
    \texttt{iplayeosgame} & \texttt{279}  & \texttt{15,803} &  \texttt{token speculator, arbitrageur}\\
\bottomrule
  \end{tabular}
\end{table}

% 这里需要修改
Account \texttt{\small newdexiofees} that holds the most token (i.e., 338 tokens) can be considered as the ``\textit{king}'' of tokens. \texttt{\small newdexiofees} is essentially an exchange initiating a large number of \texttt{\small transfer} actions (3,873,999); this is confirmed by its banner ``the first globally decentralized exchange based on EOS''\footnote{\url{https://newdex.io/}}. As for the second-rank account \texttt{\small 5lisqkvt1n2q} and third-rank account \texttt{\small iplayeosgame}, they have 284 tokens and 279 tokens, respectively. Interestingly, they also have similar \textbf{\small outdegree} and \textbf{\small invocation}. Moreover, we find that both these two accounts have frequently traded with the exchanges. Thus, we speculate that they may be token speculators who invest in EOSIO tokens to make profits. Different from \texttt{\small 5lisqkvt1n2q}, account \texttt{\small iplayeosgame} is also a gambler who frequently interacts with other gambling and gaming DApps.

\begin{figure}[t]
    \centering
    \includegraphics[width=0.46\textwidth]{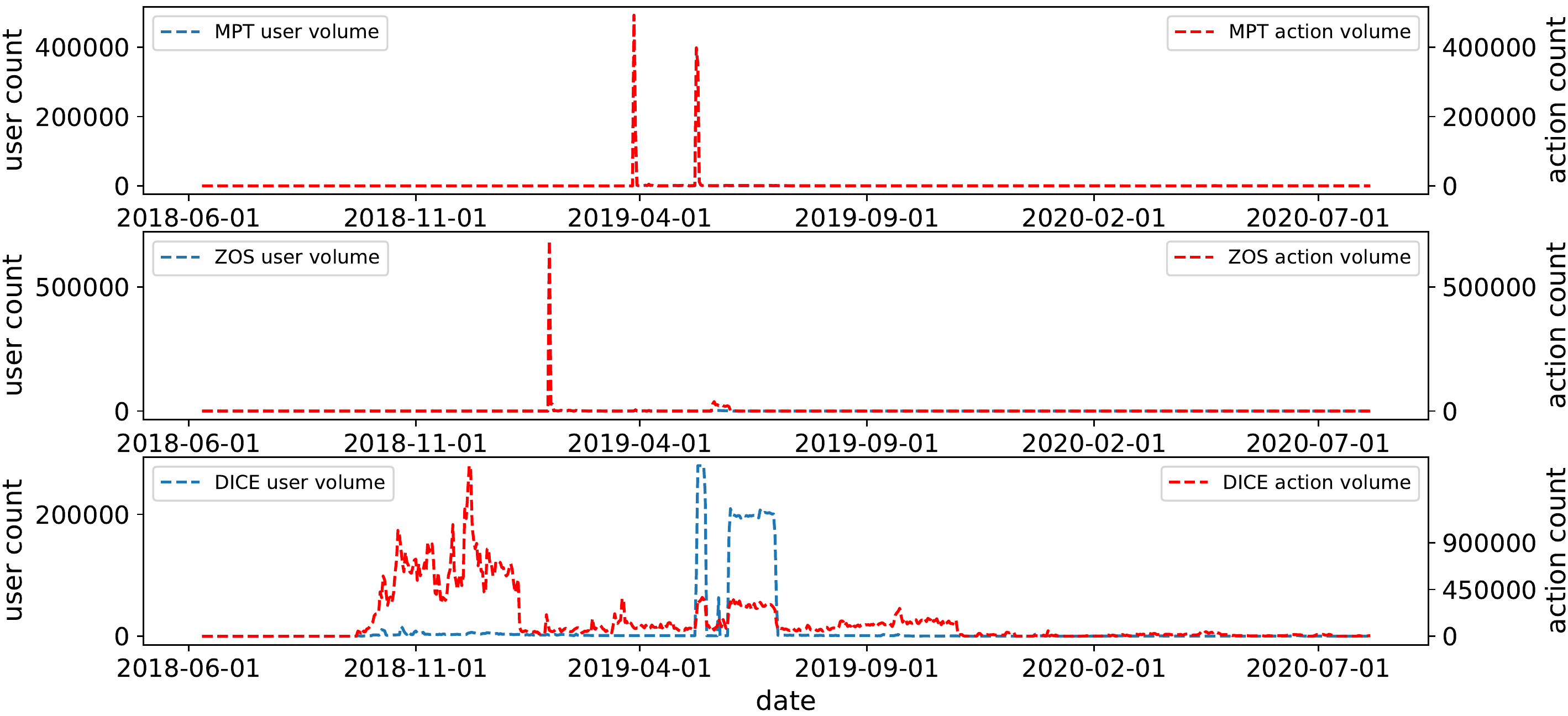}
    \vspace*{-0.25cm}
    \caption{Daily user and \texttt{\small transfer} action volume}
    \label{THG-Top-3}
\end{figure}

\textbf{Which Has the Most Holders?}
%\label{MC_WHTMH}
We then analyze the \emph{indegree distribution} of $\mathsf{THG}$. Among 5,598 tokens, 52.47\% of them only have one holder and even 78.62\% of tokens have less than 10 holders. We consider some well-known tokens that have many holders and analyze the distribution of daily participants of tokens over time. According to the number of holders, \texttt{\small MPT} is the most popular token possessed by 766,793 holders. As disclosed in DappRadar,  \texttt{\small MPT} is essentially a token for the supply chain of the metal packaging industry~\cite{MPT}. The second-rank token is \texttt{\small ZOS} (604,884 holders), which is a new token for the discount e-payment system provided by AirDropsDAC~\cite{ZOS}. The third-rank token is \texttt{\small DICE} that has 314,278 holders for \textit{BetDice}, i.e., one of the most famous gambling DApps (aforementioned in Table \ref{table3}). To explore the popularity of these tokens, we present the daily volume of users and \texttt{\small transfer} actions of top-3 tokens according to time, as shown in Fig.~\ref{THG-Top-3}. 

\begin{figure*}[h]
\centering 
\subfigure[\scriptsize $\mathsf{TTG}$]{
\centering
\includegraphics[height=1.3 in]{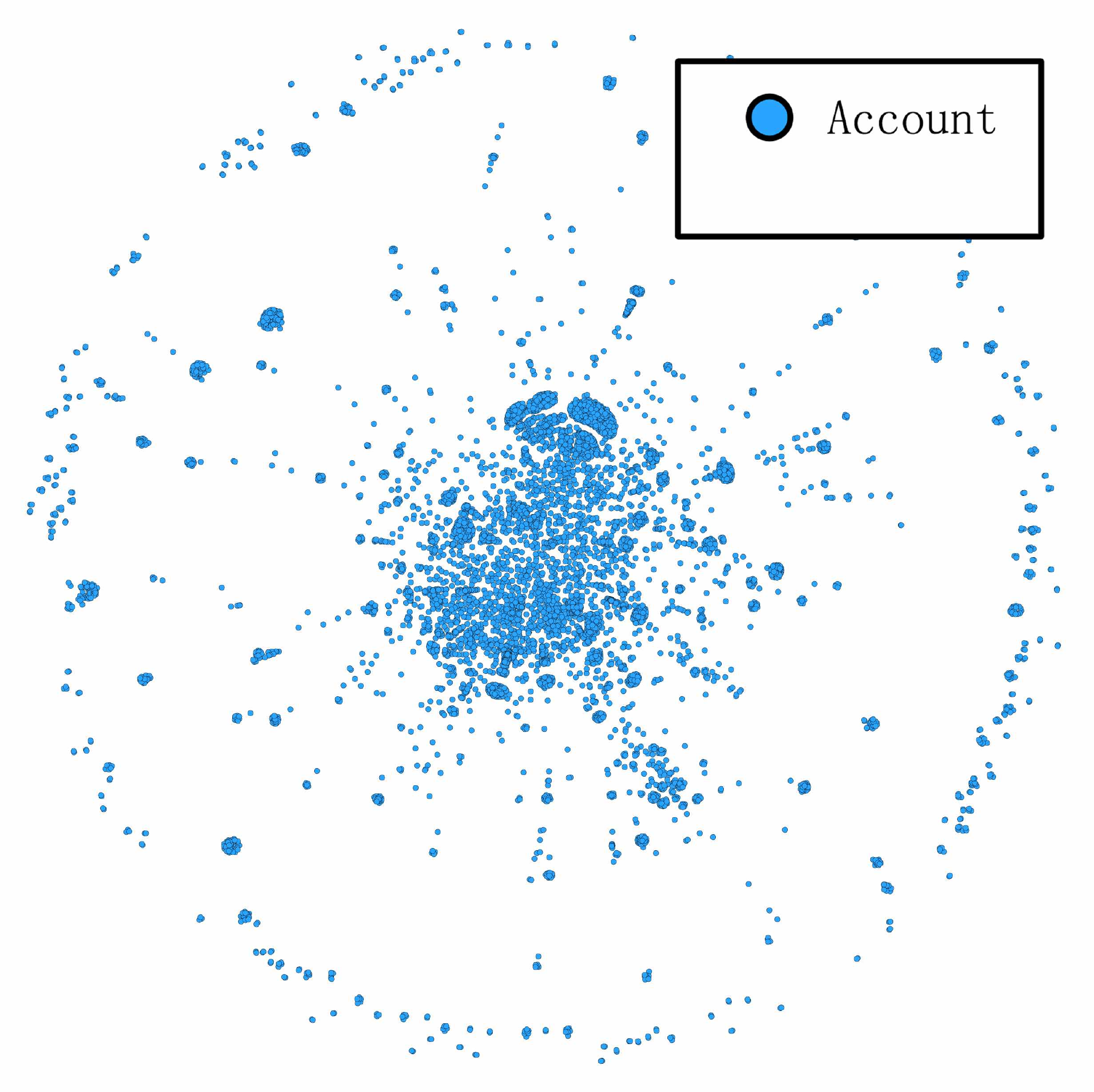}\label{TTG}
}
\quad
\subfigure[\scriptsize $\mathsf{TTG}$ in sample]{
\centering
\includegraphics[height=1.3 in]{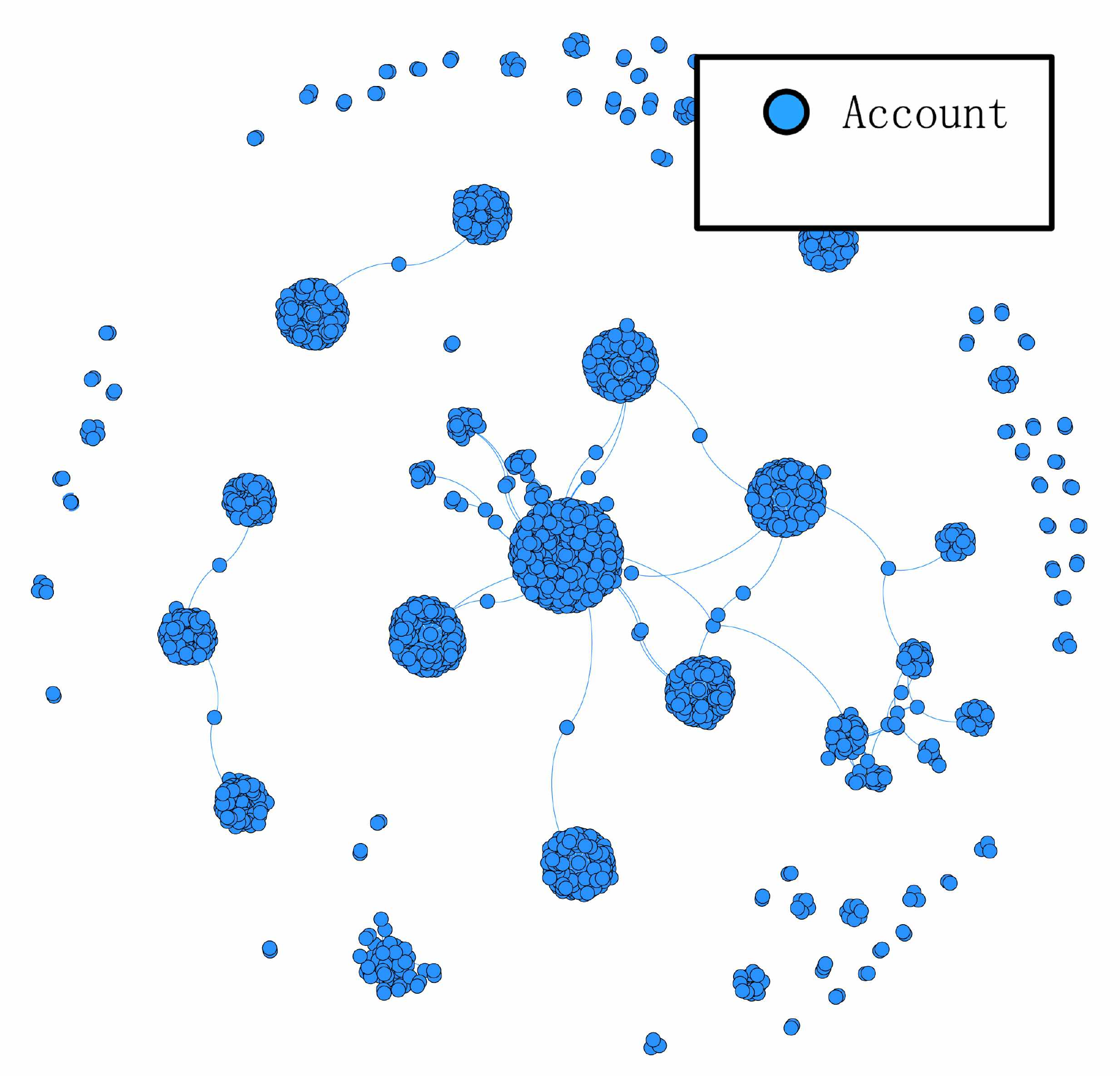}\label{sample_TTG}
}
\quad
\subfigure[\scriptsize Indegree distribution of $\mathsf{TTG}$]{
\centering
\includegraphics[height = 1.2 in]{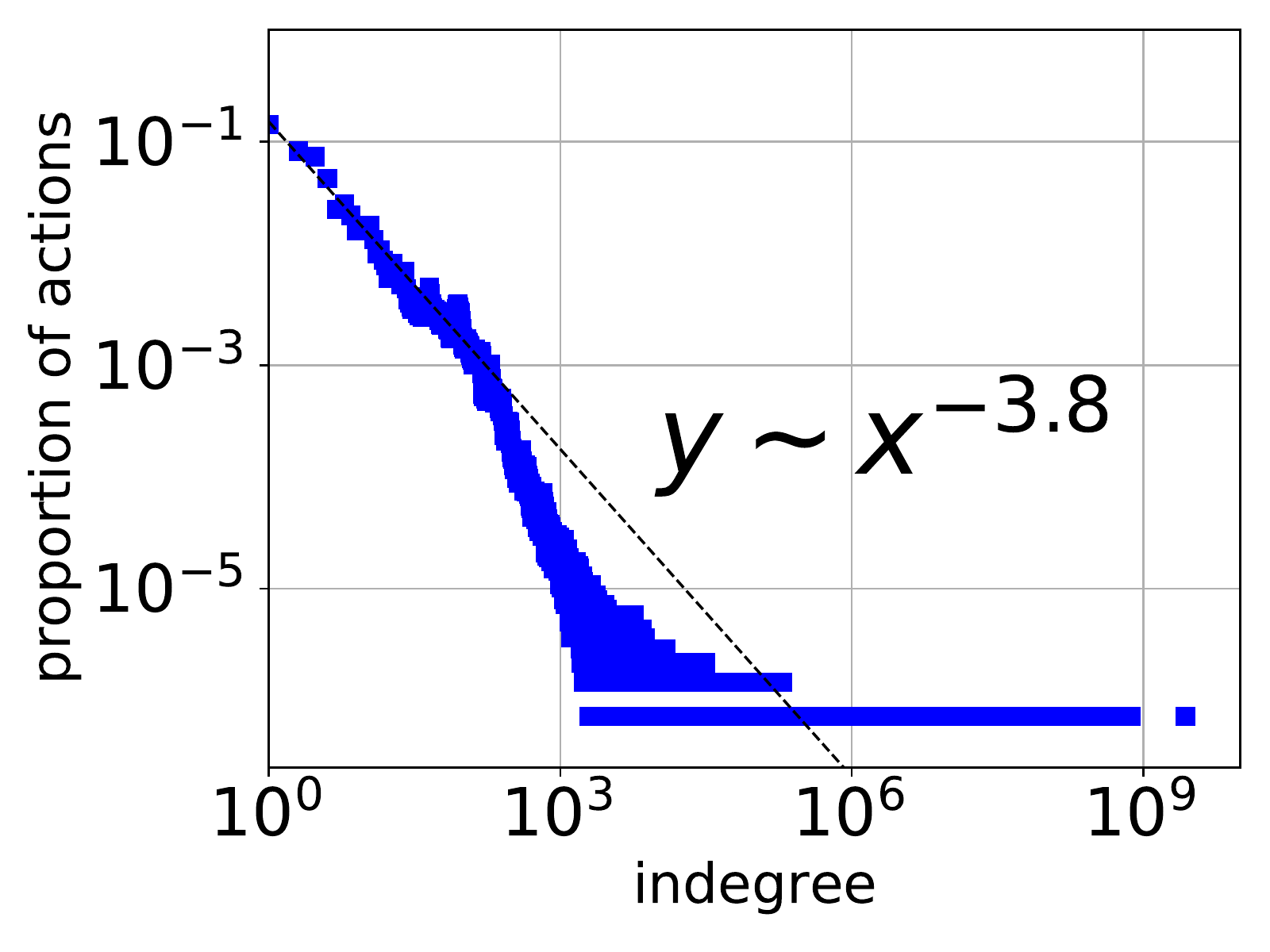}\label{in-TTG}
}
\quad
\subfigure[\scriptsize Outdegree distribution of $\mathsf{TTG}$] {
\centering
\includegraphics[height = 1.2 in]{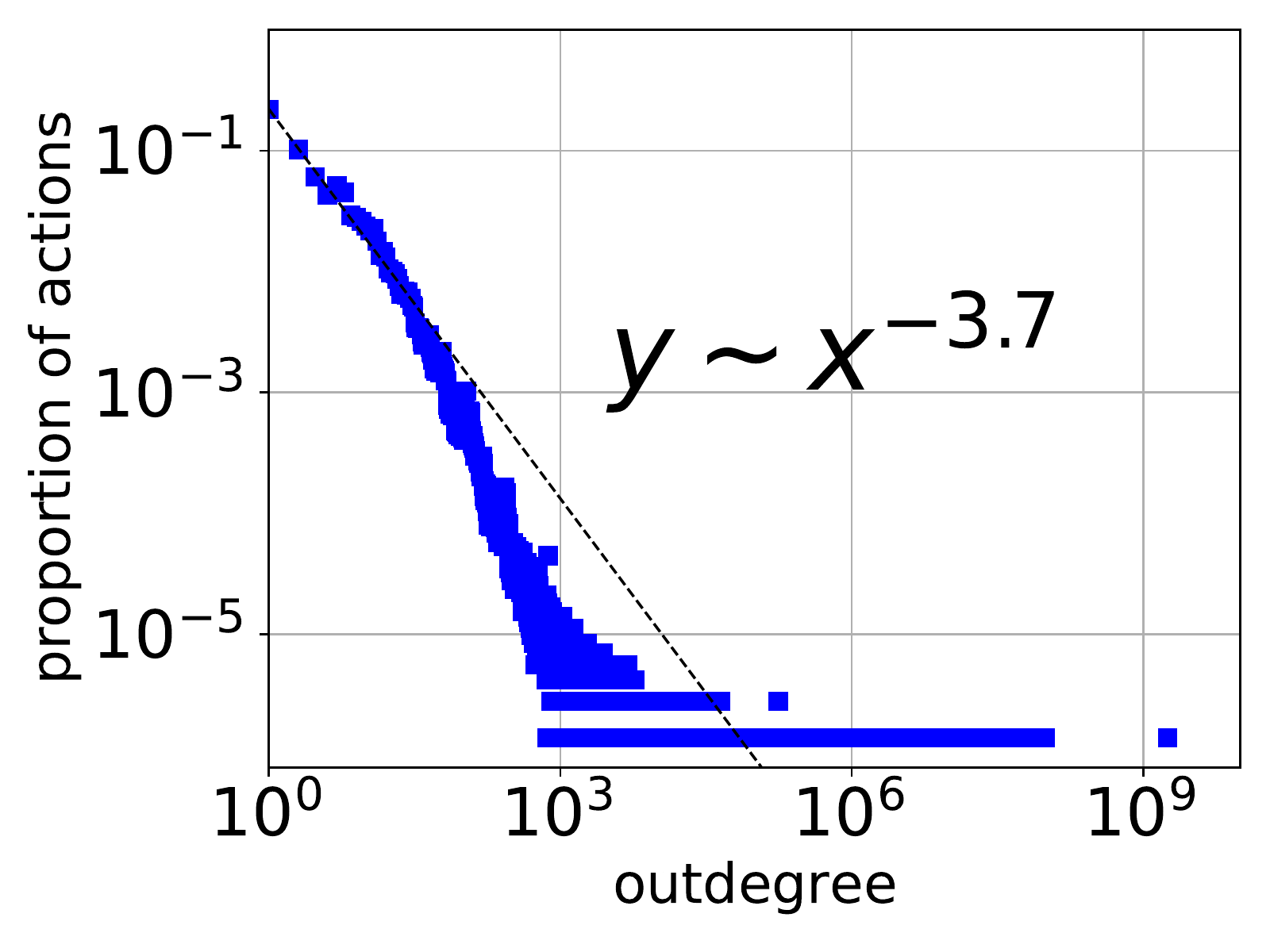}\label{out-TTG}
}
\vspace*{-0.25cm}
\caption{Visualization of Token Transfer}
\label{visualization_of_token_transfer}
\end{figure*}

Although \texttt{\small MPT} and \texttt{\small ZOS} are the top-2 tokens, their user volume and action volume have only sporadically increased (sharply). This phenomenon can be explained by the fact that these tokens (or DApps) have only received sudden attention from the public for a short time. Meanwhile, some ICO projects inject lots of fake users or fake transactions so as to arouse public attention. Section~\ref{sec:DTM} will conduct an in-depth study for this phenomenon. By contrast, there have been many participants continuously interacting with \texttt{\small DICE}, implying that gambling DApps have kept prospering in EOSIO. More interestingly, the user volume of \texttt{\small DICE} does not have the same trend as the action volume. In the early days of \texttt{\small DICE} launch, its user volume was small despite the surged action volume. This implies the importance of evaluating the token popularity from multiple perspectives.

\subsection{Comparison with Ethereum}
Despite several studies~\cite{victor2019measuring,chen2020traveling,frowis2019detecting,chen2019tokenscope} on the Ethereum token ecosystem, there are few studies on the EOSIO token system. Through the above analysis, we summarize key similarities and differences between the EOSIO and Ethereum token ecosystems.

\textbf{Similarities:} (1) \emph{The Matthew effect has been observed in both EOSIO and Ethereum} in multiple aspects like token activity, token holders, token creators. Many tokens and holders keep ``silent'' in the ecosystem. (2) \emph{Decentralized exchanges (DEX) play an important role in the token ecosystem}~\cite{DBLP:conf/www/VictorW21}. Examples include \texttt{\small newdexiofees} in EOSIO, \texttt{\small Augur} and \texttt{\small EtherDelta} in Ethereum~\cite{chen2020traveling}. Token exchange is the most active activity in the ecosystem, implying that there are a lot of arbitrage opportunities for tokens.

\textbf{Differences:} (1) \emph{The number of tokens in EOSIO is much smaller than that in Ethereum}, because the cost of deploying and maintaining a token contract in EOSIO is high (in terms of substantial resources such as CPU, RAM being staked for users). (2) \emph{One smart contact can create multiple tokens in EOSIO} although this is not allowed in Ethereum. Project parties in EOSIO are in favor of creating multiple tokens using the same contract, possibly saving the cost of token issuance. (3) \emph{Gambling and gaming are the most active activities in the EOSIO token ecosystem}. The reasons lie in the waiver of trading fees and a lower confirmation latency than Ethereum. (4) \emph{The resource-management mechanism in EOSIO is inferior to the gas mechanism in Ethereum in terms of poor security}, thereby being exploited by attackers to attack the ecosystem~\cite{lee2019spent}. For example, the DDoS attack launched by \texttt{\small eidosonecoin@EIDOS} almost exhausted CPU resources of the EOSIO \textit{mainnet}, resulting in the exceptions of other DApps or tokens.
(5) \emph{EOSIO has a much larger transaction volume than Ethereum} despite a smaller number of tokens. The major reason lies in the waiver of trading fees, which also reduces the cost of injecting fake transactions/users into DApps or tokens.

\section{Token Transfer Analysis}\label{sec:STA}
The exploratory analysis on token creators, holders, and token usage presents an exploration of the EOSIO token ecosystem. We next investigate the token transferring network and identify some abnormal trading patterns. 

\subsection{Token Transfer}\label{STA_TT}
To study the behavior characteristics of users participating in token transferring, we define the $\mathsf{TTG}$ as follows:
{
$$\mathsf{TTG}=(V_\text{tt},E_\text{tt},w),E_\text{tt}=\{(v_i,v_j,w)|v_i,v_j \in V_\text{tt}, w  \in (0,\infty)\},$$}%
where $V_\text{tt}$  is a set of the token holders and $E_\text{tt}$  is a set of edges. Each edge $(v_i,v_j,w)$ indicates that a holder $v_i$ transfers some tokens to a holder $v_j$, where $w$ is the total number of \texttt{\small transfer} actions. Hence, $\mathsf{TTG}$ is essentially a weighted directed graph. Note that we ignore the type and the amount of the transferred tokens and only count the number of \texttt{\small transfer} actions since different tokens are not comparable. As shown in Fig.~\ref{TTG}, the overall transaction network is relatively sparse while it contains some closely-connected components (i.e., trading groups). After randomly sampling 10,000 edges from Fig.~\ref{TTG}, we then reconstruct a sampled $\mathsf{TTG}$ as shown in Fig.~\ref{sample_TTG}. We further observe that many accounts are clustered together to form multiple sub-networks. To have an in-depth understanding of $\mathsf{TTG}$, we further analyze the distribution of \emph{receivers} and \emph{senders}, as depicted in Fig.~\ref{in-TTG} and Fig.~\ref{out-TTG}, respectively. In particular, the outdegree of $\mathsf{TTG}$ denotes the number of \texttt{\small transfer} actions initiated by a \emph{sender}. The indegree denotes the number of \texttt{\small transfer} actions ceased at a \emph{receiver}. The approximate degree distributions show that a large number of users keep ``silent'' in the transferring network. In addition, we find that the accounts with a large degree are often the center of the closely-connected groups as shown in both Fig.~\ref{TTG} and Fig.~\ref{sample_TTG}. We will further study these accounts and find the relationship between them.

\begin{table*}[h]
  \centering
  \setlength{\belowcaptionskip}{0.1cm}
  \caption{Top-5 Accounts of $\mathsf{TTG}$ Using Degree Centrality}
  \label{table6}
  \footnotesize
\renewcommand{\arraystretch}{0.9}
  \begin{tabular}{c|c|c|cl}
    \toprule
    \textbf{Accounts} & \textbf{Indegree} &  \textbf{Outdegree} &  \textbf{Identities}\\
    \midrule
     \texttt{eidosonecoin} & \texttt{34,137} & \cellcolor{LightCyan}\texttt{23,480,436,814} &  \texttt{Token Airdrop} \\
     \texttt{betdicegroup} & \texttt{15,582,144} & \texttt{98,538,083}  &  \texttt{BetDice, Gambling DApp}\\
     \texttt{betdicehouse} & \cellcolor{LightCyan}\texttt{58,671,515} &  \texttt{39,645,381} &  \texttt{BetDice, Gambling DApp}\\
     \texttt{betdicetoken} & \texttt{78} & \texttt{60,557,132}  &  \texttt{BetDice, Gambling DApp}\\
     \texttt{mykeypostman}  & \texttt{247,895,918} &  \texttt{28} &  \texttt{MykeyPocket, Wallet DApp}\\
\bottomrule
  \end{tabular}
\end{table*}

{\textbf{Who Transfer The Most Tokens?}}
Table~\ref{table6} shows the top-5 accounts with the largest degree. Account \texttt{\small eidosonecoin} is the issuer of token \texttt{\small EIDOS} (as mentioned in Table~\ref{table3}), which always sends \texttt{\small EIDOS} to other accounts for \texttt{\small airdrop}. Thus, account \texttt{\small eidosonecoin} has the largest outdegree of 23,480,436,814 but has a smaller indegree of 34,137. All the three accounts with a prefix ``\texttt{\small betdice-}'' belong to a gambling DApp called \textit{BetDice} though some of them have a larger indegree and some of them have a smaller indegree. This implies that they provide different functions while all working together to constitute the gambling DApp. For example, the account with a larger indegree takes stakes from gamblers while the account with a larger outdegree runs a lottery for gamblers and pays the bonus. Compared with \texttt{\small eidosonecoin}, \texttt{\small mykeypostman} has a larger indegree (247,895,918) and a smaller outdegree. We find that \texttt{\small mykeypostman} is a popular wallet DApp, which provides users with account-creation services. Since it requires purchasing RAM resources to create accounts in EOSIO, \texttt{\small mykeypostman} also requires payment from users.

\subsection{Abnormal Trading Patterns}\label{STA_STP}
\begin{figure}[h]
    \centering
    \includegraphics[width=0.4\textwidth]{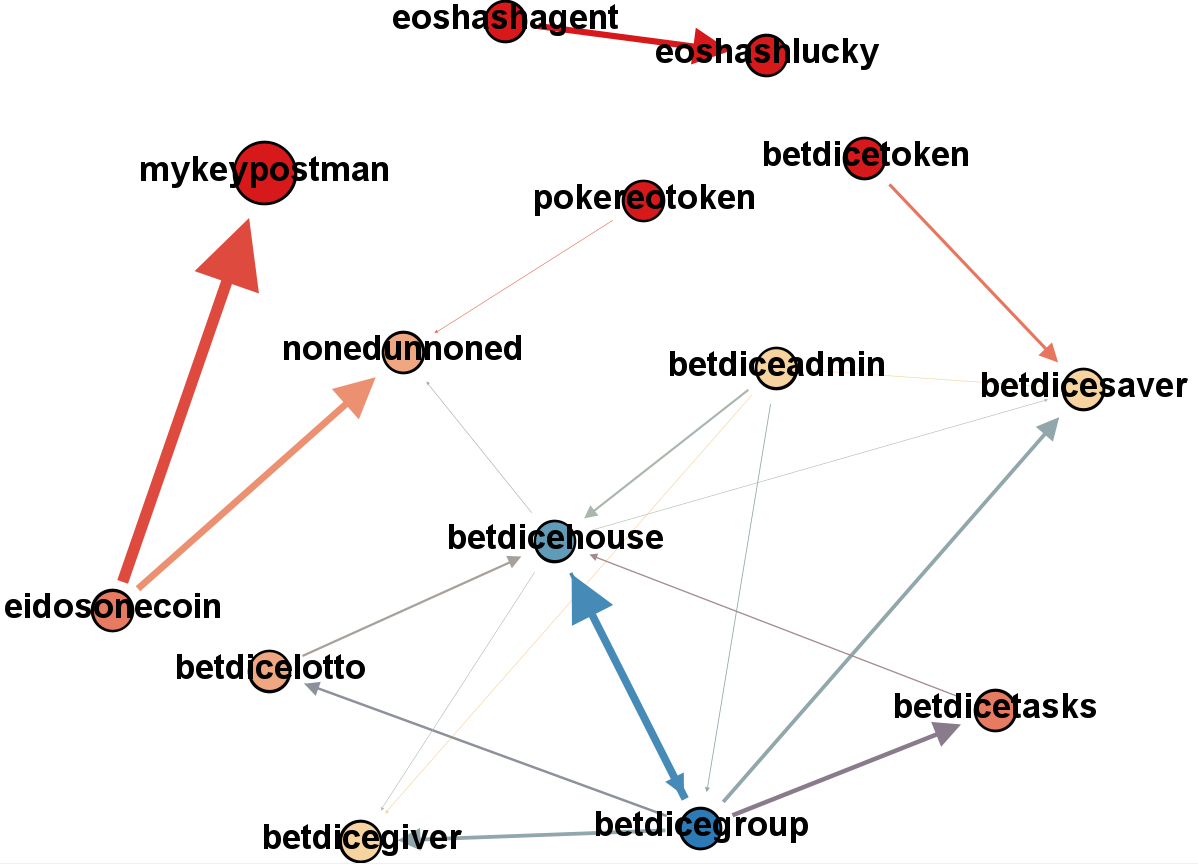}
    \vspace*{-0.25cm}
    \caption{$\mathsf{CTTG}$}
    \label{STTG}
\end{figure}

We mainly concentrate on the \textit{``center''} accounts in Fig.~\ref{sample_TTG} to find abnormal trading patterns. The main reason for focusing on ``central'' accounts lies in the \emph{relative importance} of ``central'' accounts than other ``peripheral'' accounts, where the relative importance of an account can be measured by the PageRank algorithm~\cite{pasquinelli2009google}. We first get the top-14 accounts having lots of \texttt{\small transfer} actions and define a \textit{``center''} token \texttt{\small transfer} graph ($\mathsf{CTTG}$) as follows:
{
$$\mathsf{CTTG}=(V_\text{ct},E_\text{ct},w),E=\{(v_i,v_j,w)|v_i,v_j \in V_\text{ct}, w  \in (0,\infty)\},$$}%
where $V_\text{ct}$  is a set of the top-14 accounts and $E_\text{ct}$  is a set of edges. The definition of each edge is similar to that of $\mathsf{TTG}$. The weight $w$ of each edge represents the number of \texttt{\small transfer} actions, being represented by the thickness of the edge. We can easily find some thick edges in Fig.~\ref{STTG}, which can be used to explore abnormal patterns.

According to the connection types of the nodes, we consider several abnormal patterns: 1) ``binary'' pattern, 2) ``tree'' pattern, and 3) ``grid'' pattern. As shown in Fig.~\ref{STTG}, \texttt{\small eoshashagent} frequently trades with \texttt{\small eoshashlucky} using many different types of tokens, consequently forming the ``binary'' pattern. It is abnormal for users (or investors) to trade with each other so many times, especially in a traditional financial market. Meanwhile, \texttt{\small eidosonecoin} in Fig.~\ref{STTG} often sends \texttt{\small EIDOS} to \texttt{\small mykeypostman} and \texttt{\small nonedunnoned}, which also often trades with \texttt{\small pokeneotoken}, thereby forming the ``tree'' pattern. As discussed in Section~\ref{MC_TA}, the \texttt{\small EIDOS} airdrop action can be considered as a DDoS attack. Therefore, both \texttt{\small mykeypostman} and \texttt{\small nonedunnoned} are likely to be accomplices in this attack. Moreover, all the accounts with the prefix ``\texttt{\small betdice-}'' form the ``grid'' pattern; all of them belong to a gambling DApp. It is worth mentioning that there is a thick bidirectional link between \texttt{\small betdicegroup} and \texttt{\small betdicehouse}, both of which may serve as the leaders of this gambling DApp. It is abnormal that a DApp involves so many accounts which trade with each other so frequently. So, such a trading network is likely to conduct money-laundering activities under the guise of gambling. In addition, it is also doubtful that all the accounts within the same DApp deliberately increase the transaction volume of tokens to attract huge public attention (like a scam). Further exploration of these abnormal patterns and related arguments will be carried out in our future work.

\section{Fake Token Detection}\label{sec:DTM}
The exploration on abnormal activities in the token ecosystem implies that some ICO projects and DApps may be rife with fake tokens owned by fake users to either attract sudden popularity or make exorbitant profits. 
This section aims to detect the ``\textit{fake}'' tokens.
and find out how malicious ICO projects and DApps conduct manipulation activities to make their tokens ``popular''.

\subsection{Relationship Between Accounts}\label{DTM_RBA}
\begin{figure}[t]
    \centering
    \includegraphics[width=0.3\textwidth]{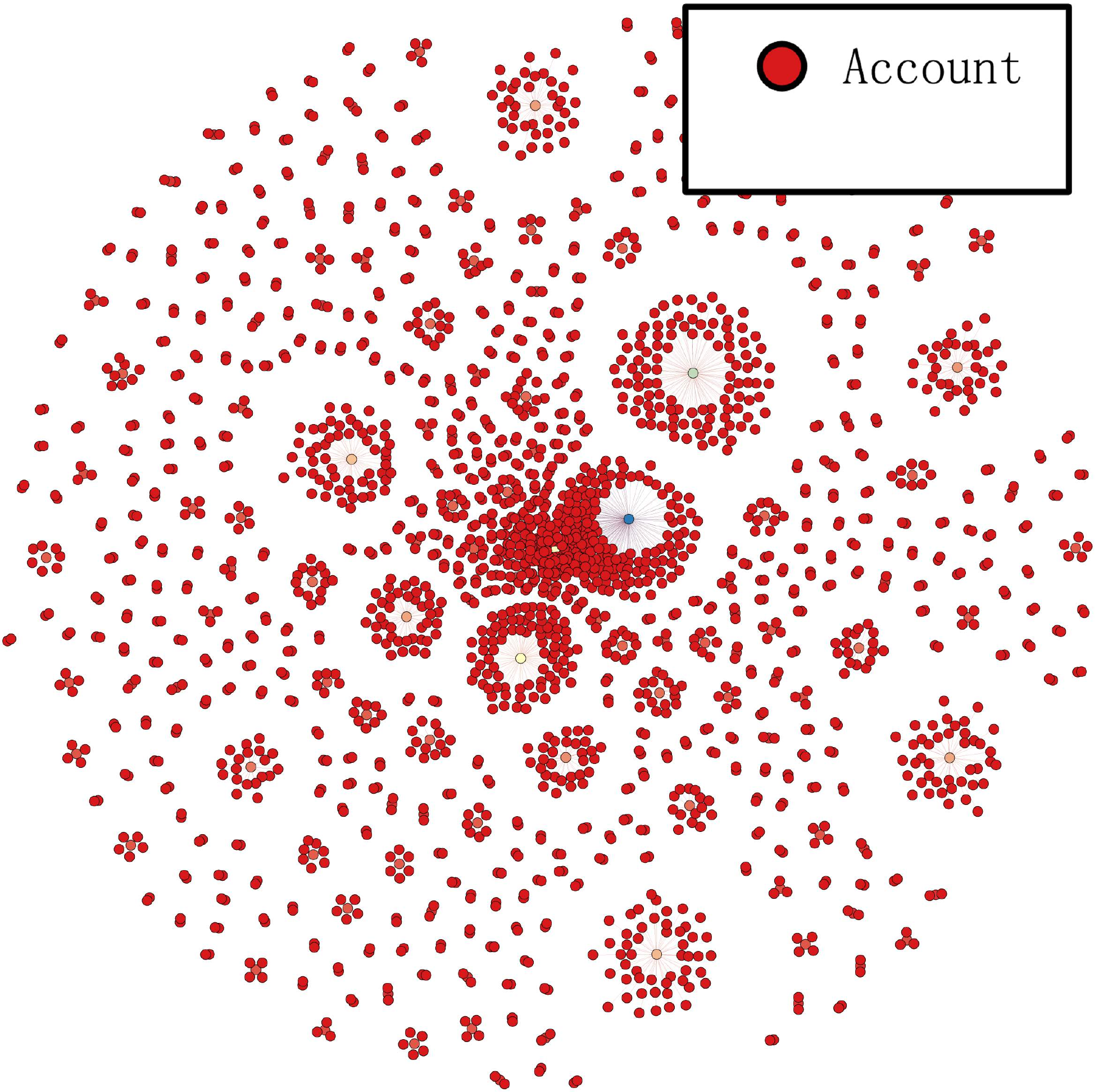}
    \vspace*{-0.25cm}
    \caption{$\mathsf{ACG}$}
    \label{ACG}
\end{figure}

We first investigate the account-creation relationship between accounts. Considering that an account \textit{Alice} in EOSIO is created by an existing account \textit{Bob}, we then regard \textit{Bob} as the parent of \textit{Alice}. To describe such a relationship, we define the account-creation graph ($\mathsf{ACG}$) as below:
{$$\mathsf{ACG}=(V_\text{ac},E_\text{ac},D),E_\text{ac}=\{(v_i,v_j,d)|v_i,v_j \in V_\text{ac}, d \in D \},$$}where $V_\text{ac}$  is a set of the accounts, $E_\text{ac}$ is a set of edges indicating the creation relationship between these accounts, and an edge $(v_i,v_j)$ represents that a parent account $v_i$ creates a child account $v_j$ on timestamp $d$. As shown in Fig.~\ref{ACG}, the result of $\mathsf{ACG}$ shows that there are a few parent accounts that have nevertheless created a large number of children accounts. When further exploring the names of these children accounts, we find a certain similarity and regularity among them. For example, many account names have the same prefix (e.g., ``\texttt{\small bnr}'', ``\texttt{\small gg}'') followed by several digits indicating their sequence number (i.e., created sequentially). These results imply that the ICO projects and DApps may adopt similar methods to create and control many fake accounts to frequently interact with their tokens, consequently flourishing their tokens.

\subsection{Algorithm}\label{DTM_A}
We then propose an algorithm to identify the ``\textit{fake}'' tokens in the ecosystem. We model the tokens and their users mainly from two dimensions. One dimension is \textbf{Average Token Transfer Number Factor (ATTNF)}, which models the number of \texttt{\small transfer} actions of users of a token. Another dimension is \textbf{Max Token Transfer Quantity Factor (MTTQF)}, which models the normalized \texttt{\small transfer} amount of users of a token. Both these two factors consider the account-creation relationship between users, which plays an important role in our algorithm. Our evaluation results also reveal a strong relationship between the token manipulator and its controlled children accounts.

{\textbf{Average Token Transfer Number Factor}.}
Considering that a token manipulator usually controls many accounts, we define the Account Control Factor ($\mathsf{ACF}$) for a token as below:
\begin{equation}
\mathsf{ACF}=\frac{\big|\{\text{holder}_i|i=1,2,...,n\}\big|}{\big|\{\text{parent}_j|j=1,2,...,m\}\big|} \ \text{and} \ m \leq  n,
\end{equation} where $\text{holder}_i$ represents a unique account $i$ who transfers the token and $\text{parent}_j$ represents a parent account of holders in the set $\{\text{holder}_i|i=1,2,...,n\}$. For convenience, $\{\text{parent}_j|j=1,2,...,m\}$ and $\{\text{holder}_i|i=1,2,...,n\}$ are abbreviated to $\mathbb{P}$ and $\mathbb{H}$, respectively. $\mathsf{ACF}$ is the ratio of the size of $\mathbb{H}$ to that of $\mathbb{P}$.
% $|\bigcup_1^{n} \text{holder}_i|$ and $|\bigcup_1^{k} \text{parent}_i|$ represents the number of holders and their parent accounts, respectively. 
If a token is only transferred by the accounts who have the same parent, it is quite possible that the parent creates a large number of fake children accounts to manipulate transactions, leading to a large $\mathsf{ACF}$.

However, it is not enough to only consider the relationship between the parent and its children accounts, because token manipulators who have created lots of children accounts often have an aim to conduct \texttt{\small transfer} actions including many fake transactions. Hence, we define another factor, Action Number Factor ($\mathsf{ANF}$) to further model \texttt{\small transfer} actions. $\mathsf{ANF}$ is defined as follows:
\begin{equation}
\mathsf{ANF}^{T_k}_{\text{holder}_i} = \frac{\mathsf{NUMBER}(\text{holder}_i,T_k)}{\mathsf{NUMBER}(\text{holder}_i,\{T_k|k=1,2,...,z\})},
\label{eq:anf}
\end{equation} where ${\scriptstyle \mathsf{NUMBER}(\text{holder}_i,T_k)}$ represents the number of the \texttt{\small transfer} actions on the token $T_k$ initiated by the account $\text{holder}_i$. Set $\{T_k|k=1,2,...,z\}$ represents all tokens held by ${\text{holder}_i}$ and $\scriptstyle \mathsf{NUMBER}(\text{holder}_i,\{T_k|k=1,2,...,z\})$ denotes the number of all the \texttt{\small transfer} actions of $\text{holder}_i$ on all tokens he/she holds. In other words, if $\mathsf{ANF}^{T_k}_{\text{holder}_i}=1$, it implies that $\text{holder}_i$ is created only for interacting with token $T_k$. For a token ${T_k}$, we get its Total Action Number Factor ($\mathsf{TANF}$) across all its holders as follows:
\begin{equation}
\mathsf{TANF}=\mathsf{ANF}^{T_k}_{\text{holder}_1}+\mathsf{ANF}^{T_k}_{\text{holder}_2}+...+\mathsf{ANF}^{T_k}_{\text{holder}_n}.
\end{equation}
To a certain extent, $\mathsf{TANF}$ reflects the ``loyalty'' of users to a token. If $\mathsf{TANF}$ is very large, it means that almost all holders of a token only hold and transfer this token forever. It is possible that these accounts are manipulated to increase the transaction volume of the token. $\mathsf{TANF}$ that is only evaluated from the behaviors of token holders does not consider account-creation relationships like $\mathsf{ACF}$. Thus, we should consider both $\mathsf{ACF}$ and $\mathsf{TANF}$ together for each token. One naive method is dividing $\mathsf{TANF}$ by $|\mathbb{P}|$($|\mathbb{P}|$=$|\{\text{parent}_j|j=1,2,...,m\}|$), i.e., ${\mathsf{TANF}}/{|\mathbb{P}|}$. The smaller $|\mathbb{P}|$ leads to the larger ${\mathsf{TANF}}/{|\mathbb{P}|}$, implying that this token may be controlled by a few parent accounts. However, this naive method is not optimal due to the following reason. In EOSIO, there are several wallet DApps that help common users create a large number of accounts. When a token is really popular, many (but not all) accounts whose parent is a wallet DApp will follow to participate ($|\mathbb{P}|$ may be small). This may mistakenly cause a large value of ${\mathsf{TANF}}/{|\mathbb{P}|}$, consequently leading to some false positives.

To address this issue, we model the behaviors of each parent (using $M_{\text{parent}_j}$) instead of simply counting the number. We finally define the $\mathsf{ATTNF}$ for a token as follows:
\begin{equation}
    \mathsf{ATTNF} = \frac{\mathsf{TANF}}{\sum_{j=1}^m M_{\text{parent}_j}},
\label{eq:aacf}
\end{equation}
where $M_{\text{parent}_j}=\frac{|\{\text{child}_i|i=1,2,...,N\}|}{|\{\text{holder}_i|i=1,2,...,n\}|}$ for each parent account. The set $\{\text{child}_i|i=1,2,...,N\}$ (abbreviated as $\mathbb{C}$) is the total accounts created by $\text{parent}_j$ and $\{\text{holder}_i|i=1,2,...,n\}$ denotes the accounts who are created by $\text{parent}_j$ and hold the token. Thus, we have the set $\mathbb{C} \subseteq \mathbb{H}$ and $n \leq N$. For a token, we calculate all its $M_{\text{parent}_j}$ by dividing $|\mathbb{C}|$  by $|\mathbb{H}|$ and add them up to get $\sum_{j=1}^m M_{\text{parent}_j}$. To create a fake token, the manipulator generally exploits almost all of its children accounts to initiate lots of \texttt{\small transfer} actions in a short time. So its $M_{\text{parent}_j}$ is nearly equal to 1. On the contrary, the behavior of children accounts of a wallet DApp is more scattered and only partial children accounts interact with the token. Thus, its $M_{\text{parent}_j}$ is greater than 1. The sum 
$\sum_{j=1}^m M_{\text{parent}_j}$ is still large when many children accounts of a wallet DApp (whose $M_{\text{parent}_j}$ is relatively large) participate in a real popular token, leading to a relatively small $\mathsf{ATTNF}$ and alleviating the problem of false positives. Meanwhile, if a token is ``\textit{fake}'', almost each $M_{\text{parent}_j}$ is nearly equal to 1 and $\sum_{j=1}^m M_{\text{parent}_j}$ is relatively small, leading to a large $\mathsf{ATTNF}$. To this end, $\mathsf{ATTNF}$ is an important indicator to measure whether a token is ``\textit{fake}''. The larger the $\mathsf{ATTNF}$ is, the more likely the token is ``\textit{fake}''.

{\textbf{Max Token Transfer Quantity Factor}.}
In addition to the number of transactions and the number of holders, the total amount of a token being transferred has also attracted public attention. Thus, we also consider the \texttt{\small transfer} quantity (i.e., the amount of \texttt{\small transfer} actions in EOSIO). To measure it, we first divide holders of a token into multiple account groups, each of which has the same parent. We denote such an account group by $\{\text{holder}_i|i=x,x+1,...,y\}$. We then define the Token Transfer Quantity Factor ($\mathsf{TTQF}$) as follows:
\begin{equation}
\small
\mathsf{TTQF} = \sum^y_x \mathsf{Qua}_i = \sum^y_x\frac{\mathsf{Qua}(\text{holder}_i,T_k)}{\mathsf{Qua}(\text{holder}_i,\{T_k|k=1,2,...,z\})} \ \text{and} \ i \in [x, y],
\label{eq:qua4}
\end{equation}
% where the parent of the account group $\{\text{holder}_i|i=x,x+1,...,y\}$ is the same 
% \begin{equation}\footnotesize ACF = \frac{|\{\text{holder}_i|i=x,x+1,...,y\}|}{|\{\text{parent}_j|j=1,2,...,k\}|} = |\{\text{holder}_i|i=x,x+1,...,y\}| \label{eq:acf}\end{equation} 
where \begin{equation}\mathsf{Qua}(\text{holder}_i,T_k)=\frac{\text{total transfer quantity of } \text{holder}_i \text{ on  token } T_k}{\text{issue quantity of token } T_k}\label{eq:qua}\end{equation}
and
\begin{equation}
\mathsf{Qua}(\text{holder}_i,\{T_k|k=1,2,...,z\})=\sum_{k=1}^z \mathsf{Qua}(\text{holder}_i,T_k)
\label{eq:qua6}
\end{equation}
% Eq.~(\ref{eq:acf}) is a constraint condition for $\mathsf{TTQF}$, indicating that $\mathsf{TTQF}$ is calculated for an account group $\{\text{holder}_i|i=x,x+1,...,y\}$ whose parent is the same, that is $|\{\text{parent}_j|j=1,2,...,k\}|$. 
In Eq.~(\ref{eq:qua}), $\mathsf{Qua}(\text{holder}_i,T_k)$ denotes the ratio of the transferring quantity of an account (in an account group) to the total issuance quantity of a token. Since the total issuance of different tokens is not the same, it is necessary to normalize $\mathsf{Qua}(\text{holder}_i,T_k)$. Similar to the definition of $\mathsf{ANF}$ in Eq.~(\ref{eq:anf}), the set $\{T_k|k=1,2,...,z\}$ in Eq.~(\ref{eq:qua4}) and Eq.~\eqref{eq:qua6} represents all tokens held by $\text{holder}_i$.
% In the definition of $\mathsf{TTQF}$, $\mathsf{Qua}(\text{holder}_i,\bigcup_1^{z} T_j)$ is equal to $\sum_{j=1}^z \mathsf{Qua}(\text{holder}_i,T_j)$, in which $\bigcup_1^{z} T_j$ represents all tokens held by $\text{holder}_i$. 
If $\mathsf{Qua}_i = \frac{\mathsf{Qua}(\text{holder}_i,T_k)}{\mathsf{Qua}(\text{holder}_i,\{T_k|k=1,2,...,z\})} = 1$, it means that $\text{holder}_i$ only holds one token and transfers this token. We finally add up all $\mathsf{Qua}_i$ of each $\text{holder}_i$ to get the  $\mathsf{TTQF}$ for an account group. If $\mathsf{TTQF}=|x-y| + 1$ for a token, it means that this account group only holds and transfers this token. Thus, it may be a suspicious group of the token controlled by the manipulator. A large value of $\mathsf{TTQF}$ means that this group that has a large scale almost only interacts with this token. Regarding a token, there are generally multiple account groups. We define the $\mathsf{MTTQF}$ for a token as:
\begin{equation}
    \mathsf{MTTQF}=\max(\mathsf{TTQF}_1,\mathsf{TTQF}_2,...,\mathsf{TTQF}_q).
\label{eq:mtqf}
\end{equation}
A token that has a larger $\mathsf{MTTQF}$ also has a higher possibility of being manipulated. As another important indicator considering both account-creation relationship and transfer amount, $\mathsf{MTTQF}$ is helpful for finding fake accounts and the manipulator behind them. 

\begin{algorithm}
\DrawBox[draw=LightPink,fill=LightPink!30]{a}{b}
\DrawBox[draw=LightBlue,fill=LightBlue!30]{c}{d}
\footnotesize
 \caption{Search maximum $\mathsf{ATTNF}$ or $\mathsf{MTTQF}$}
 \label{alo}
 \begin{algorithmic}[1]
 \Require \texttt{\footnotesize Actions[<sender,token,quantity,sender\_parent>]}, Window Size $W$, Pieces $P$, Flag $F$
 \Ensure maximum $\mathsf{ATTNF}$ or $\mathsf{MTTQF}$ 
 \State $\mathsf{arr} \gets [], \ \text{piece}_\text{size} \gets \frac{W}{P}, \ \text{piece}_\text{count} \gets \frac{|\texttt{Actions}|}{\text{piece}_\text{size}}$ 
  \For {$i = 0 \to (\text{piece}_\text{count} - 1)$} \tikzmark{a}
   \State $\text{piece}_\text{start} \gets i \times \text{piece}_\text{size},\ \text{piece}_\text{end} \gets (i + 1) \times \text{piece}_\text{size}$
   \State $\text{res} \gets \mathsf{ATTNF}\_OR\_\mathsf{MTTQF}(\texttt{Actions}[\text{piece}_\text{start} : \text{piece}_\text{end}],\ F)$
   \State push $\text{res}$ into $\mathsf{arr}$ 
 \EndFor\tikzmark{b}
 \State $\text{sum}_{\max} \gets 0$
 \For {$i = 0 \to (P - 1)$}
    \State $\text{sum}_{\max} \gets \text{sum}_{\max} + \mathsf{arr}[i]$
 \EndFor
 \State $\text{temp} \gets \text{sum}_{\max}, \ \text{index}_{\max} \gets 0$\tikzmark{c}
 \For {$i = P \to (|\mathsf{arr}| - 1)$}
    \State $\text{temp} \gets \text{temp} + \mathsf{arr}[i] - \mathsf{arr}[i - P]$
    \If {$\text{temp} > \text{sum}_{\max}$} \Comment{{\scriptsize calculate the maximum of continuous $P$ pieces}}
      \State $\text{sum}_{\max} \gets \text{temp}, \text{index}_{\max} \gets i - P + 1$
    \EndIf
 \EndFor\tikzmark{d}
 \State $\text{window}_\text{start} \gets \text{index}_{\max} \times \text{piece}_\text{size}, \ \text{window}_\text{end} \gets \text{window}_\text{start} + W$ \Comment{{\scriptsize  $\text{index}_{\max}$ to calculate window range}}
 \State \Return $\mathsf{ATTNF}\_OR\_\mathsf{MTTQF}(\texttt{Actions}[\text{window}_\text{start} : \text{window}_\text{end}],\ F)$
 \end{algorithmic}
 \end{algorithm}
%  \setlength{\floatsep}{0.01cm}
 % (with maximum value)

{\textbf{Search For Maximum $\mathsf{ATTNF}$ And $\mathsf{MTTQF}$}.}
For most token manipulators, they always deluge their tokens with fake users and fake transactions. There is often a surge of transactions within a short time. Once enough popularity (or investments) has been received, the volume of transactions quickly slumps. It is challenging to capture this phenomenon if we only calculate $\mathsf{ATTNF}$ and $\mathsf{MTTQF}$ using all historical records, thereby missing many ``\textit{fake}'' tokens. Addressing this issue requires selecting an appropriate window to include the maximum value of $\mathsf{ATTNF}$ or $\mathsf{MTTQF}$. To this end, we propose Algorithm~\ref{alo}, where the input includes \texttt{\small Actions}, Window Size $W$, Pieces $P$, and Flag $F$. \texttt{\small Actions} contain all the \texttt{\small transfer} actions of a token and also the parent information of \emph{senders}. $W$ is the number of actions in a window and $F$ indicates whether looking for $\mathsf{ATTNF}$ or $\mathsf{MTTQF}$. We first divide each window into $P$ pieces, each of which is a small window with size $\text{piece}_\text{size}$. \texttt{\small Actions} are divided into $\text{piece}_\text{count}$ pieces. We then calculate $\mathsf{ATTNF}$ or $\mathsf{MTTQF}$ of each small window through sliding one piece and save them into array $\mathsf{arr}$ (\hlpink{lines 2 to 6}). We next adopt the greedy strategy to obtain the maximum sum of the continuous $P$ pieces as well as the corresponding index $\text{index}_{\max}$ (\hlblue{lines~11 to~17}). Thus, we regard $\text{index}_{\max}$ as the target to seek for a window and calculate its $\mathsf{ATTNF}$ or $\mathsf{MTTQF}$ (lines~\textbf{18} to~\textbf{19}). The sliding mode based on the small window can find a larger $\mathsf{ATTNF}$ or $\mathsf{MTTQF}$, improving the accuracy of Algorithm~\ref{alo}. Note that $\mathsf{ATTNF}\_OR\_\mathsf{MTTQF}(\texttt{\small Actions}[x:y],F)$ is given in both Eq. (\ref{eq:aacf}) and Eq. (\ref{eq:mtqf}).
%while flag $F$ indicates which one to use.

\subsection{Evaluation Results}\label{DTM_E}
\begin{figure}[h]
    \centering
    \includegraphics[width=0.40\textwidth]{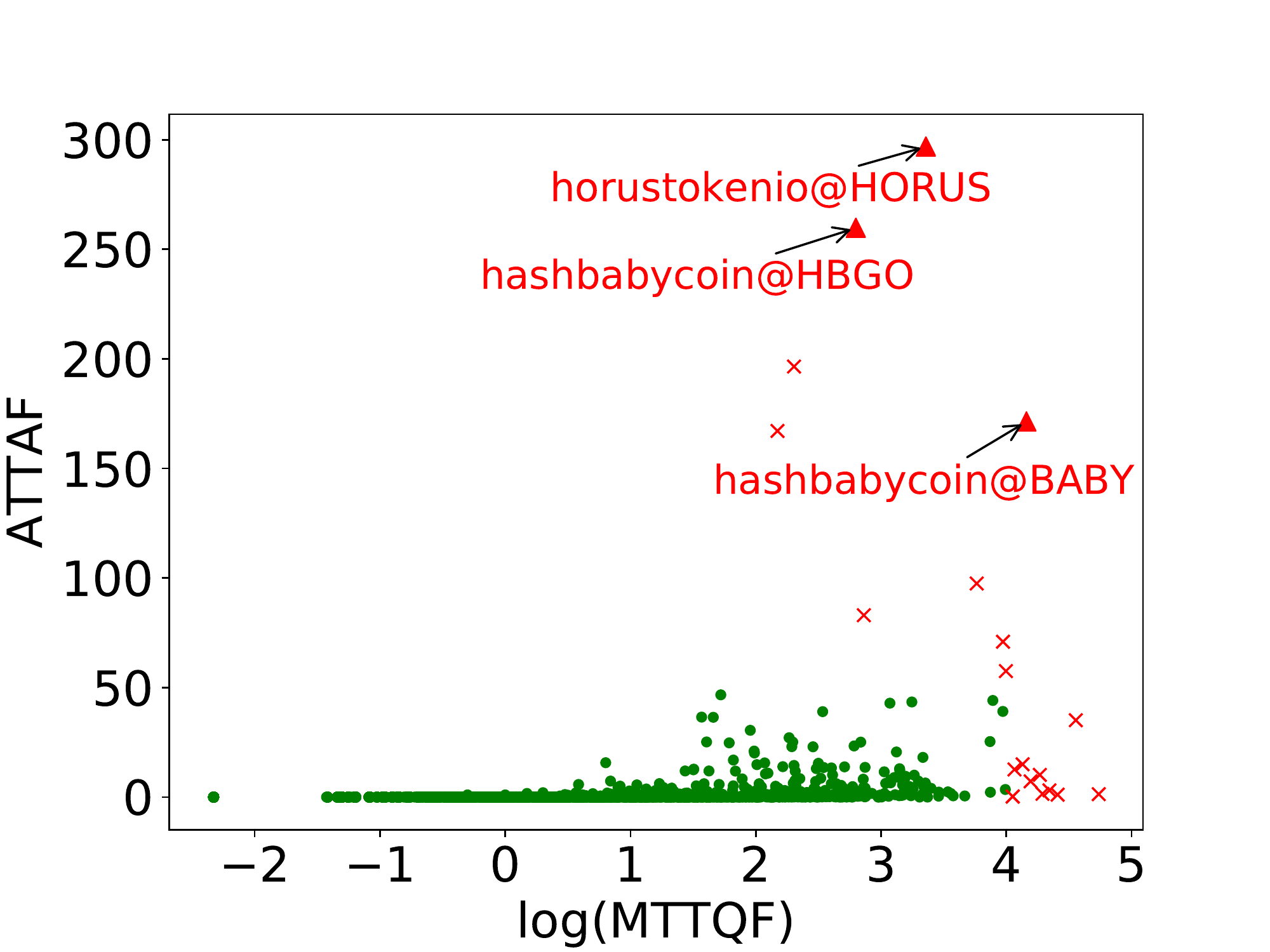}
    \vspace*{-0.25cm}
    \caption{Visualization of normal and suspicious tokens}
    \label{detection}
\end{figure}

We implement Algorithm~\ref{alo} with Python. In our experiment, we set $W = 100,000$ and $P = 10$. After calculating the maximum $\mathsf{ATTNF}$ and $\mathsf{MTTQF}$ for each token, we finally visualize the distribution of these two factors, as shown in Fig.~\ref{detection}, where we adopt the logarithmic form of $\mathsf{MTTQF}$ because of its large variance.

\begin{figure*}[ht]
    \centering
    \includegraphics[width=0.8\textwidth]{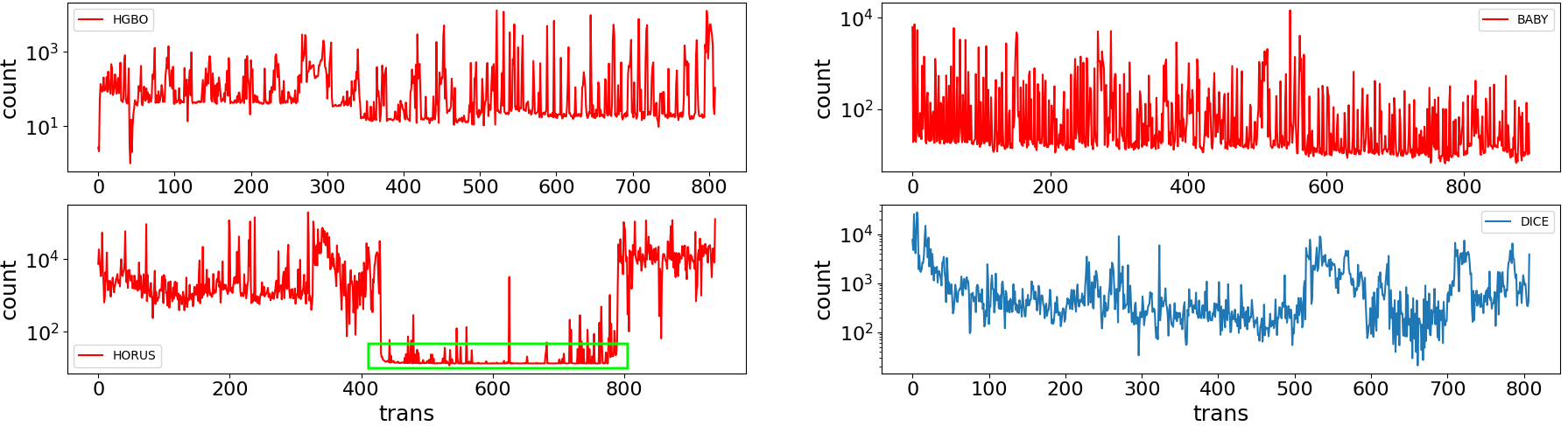}
    \caption{Visualization of normal and suspicious tokens}
    \vspace*{-0.25cm}
    \label{HORUS}
\end{figure*}

We mark suspicious tokens in red as their ${\mathsf{\small ATTNF}}$ or ${\mathsf{\small MTTQF}}$ is at a high level ($\mathsf{\small ATTNF} > 50\ or \  \mathsf{\small MTTQF} > 10,000$). In particular, we select the top-3 tokens (with large $\mathsf{ATTNF} \times \mathsf{MTTQF}$ products): \texttt{\small HBGO}, \texttt{\small BABY}, and \texttt{\small HORUS}. We then focus on these three tokens and investigate the manipulation behaviors of masterminds as well as fake transactions. To achieve this goal, we select a normal token \texttt{\small DICE} and compare it with these three tokens. We randomly sample the \texttt{\small transfer} actions of these four tokens and analyze the quantity distribution of each action. As shown in Fig.~\ref{HORUS}, the distribution of \texttt{\small DICE} presents an irregular fluctuation while the top-3 tokens periodically have high volumes of \texttt{\small transfer} actions with a relatively-fixed quantity. Meanwhile, these \texttt{\small transfer} actions have been submitted in a short time. Further, it can be observed from the green box in Fig.~\ref{HORUS} that \texttt{\small HORUS} has a large number of \texttt{\small transfer} actions with 20.00 \texttt{\small HORUS}. We next explore evidences on the existence of fake users or fake transactions of these three tokens.

\textbf{hashbabycoin@HBGO:} \texttt{\small HBGO} token that has served as a famous pornographic DApp was created by \texttt{\small pornhashbaby} through the contract \texttt{\small hashbabycoin}. The work~\cite{huang2020characterizing} has reported that \texttt{\small pornhashbaby} is the controller who has created eight groups of bot-like accounts. Each group of them has hundreds to thousands of accounts. It is quite possible that \texttt{\small HBGO} has been controlled by \texttt{\small pornhashbaby}. When scanning the \texttt{\small transfer} actions, we find that \texttt{\small pornhashbaby} usually sends 1.0000 \texttt{\small HBGO} to the accounts when being registered as users. Most of the names of these accounts have a common prefix like ``k'', ``z'', ``gi'', and ``gg''. Meanwhile, these names are sorted according to alphabet letters (a-z) or decimal digits (1-9). In addition, the transfer amount of most actions is fixed in a period (e.g., 11, 47). Locating the parent of the accounts, we find that a large number of accounts involved \texttt{\small HBGO} were created by \texttt{\small moneyloveyou}, \texttt{\small eosbank54321}, and \texttt{\small greedysogood}. These accounts may be accomplices who assist \texttt{\small pornhashbaby} to manipulate the token \texttt{\small HBGO}. More interestingly, all of the three accounts have been created by \texttt{\small Meetone}, another well-known DApp.

% Besides the evidence of ``\textbf{hashbabycoin@HBGO}'' given in \S~\ref{DTM_E}, we also present additional evidences on other ``\textit{fake}'' tokens as follows.

{\textbf{hashbabycoin@BABY:} \texttt{\small BABY}}
The same as the \texttt{\small HBGO}, \texttt{\small BABY} is another token created through the token contract \texttt{\small hashbabycoin} by \texttt{\small pornhashbaby}. We observe some similar phenomena on \texttt{\small BABY}. For example, there are a large number of \texttt{\small transfer} actions done by \texttt{\small pornhashbaby}, which sends 11.0000 \texttt{\small BABY} to other accounts.  Among them, 41,956 accounts that are prefixed with ``bnr'' have all been created by \textit{walletbancor}. These accounts periodically interact with \texttt{\small BABY}. It is shown in the top two sub-figures of Fig.~\ref{HORUS} that both \texttt{\small HBGO} and \texttt{\small BABY} have a similar quantity distribution with a periodical trend. Surprisingly, there are 7,173,443 \texttt{\small transfer} actions involved in the accounts with the prefix ``bnr*'', accounting for 43.25\% of the total transaction volume of \texttt{\small BABY}. 

{\textbf{horustokenio@HORUS:} \texttt{\small HORUS}}
The contract \texttt{\small horustokenio} (\footnote{\url{https://horuspay.io/}}) represents an entity called \textit{HorusPay} mainly used for companies to exchange private encrypted data. \texttt{\small HORUS} is one of the tokens created by \texttt{\small horustokenio}. After analyzing its action records, we find some abnormal \texttt{\small transfer} actions. For example, nearly 9,000 actions involve the accounts named ``g*ge" or "h*ge" and \texttt{\small chainceoneos} from July 17, 2018 to Aug. 13, 2018. Meanwhile, \texttt{\small chainceoneos} transfers \texttt{\small HORUS} to \texttt{\small chainceout11} several times, each \texttt{\small transfer} action is associated with a large amount of \texttt{\small HORUS} tokens (from 300,000.0000 to 15,845,927.6564). More interestingly, we observe that \texttt{\small chainceout11} frequently interacts with the accounts named ``g*ge'' or ``h*ge'' and transfers \texttt{\small HORUS} to them. It seems that these accounts have formed a closed loop between \texttt{\small chainceoneos} and \texttt{\small chainceout11}. It is reasonable to suspect that it is a manipulation of \texttt{\small HORUS}, attempting to make \texttt{\small HORUS} be ``popular''.

\section{Related Work}\label{sec:RW}
\subsection{EOSIO Analysis}
There are a number of studies on blockchain data analytics on Ethereum and Bitcoin~\cite{ron2013quantitative,reid2013analysis,kondor2014rich,chen2018understanding,wu2020phishers,torres2019art,hu2021transaction,LIU2022158,9614336}. Most of them focus on user behaviors, cryptocurrency flows, and scams of blockchains. Despite the popularity of EOSIO, there are a few systematic studies on the EOSIO ecosystem. XBlock-EOS\cite{zheng2020xblock} provides an efficient method of data extraction and exploration on the EOSIO blockchain data. Meanwhile, some recent studies characterize different types of activities in EOSIO (such as money transfer and contract invocation) and attempt to identify some bots and fraudulent activities~\cite{huang2020characterizing,zhao2020exploring}. Moreover, other studies focus on detecting vulnerable EOSIO contracts~\cite{huang2020eosfuzzer,quan2019evulhunter,he2020security}. Further, studies \cite{lee2019push,lee2019spent} find design defects of the EOSIO framework, which can be exploited by attackers. However, most of the existing studies either focus on the visualization of EOSIO's various activities or identify security vulnerabilities of EOSIO. There is no work to explore EOSIO from the cryptocurrency ecosystem perspective. It is critical for EOSIO cryptocurrency stakeholders to fully understand the EOSIO token ecosystem. This paper aims to bridge this gap by conducting a comprehensive analysis on the EOSIO token ecosystem.

\subsection{Token Analysis}
In recent years, the prosperity of ICOs has brought immeasurable values to blockchains, such as Ethereum and EOSIO. As the crucial component in value-transferring process of blockchains, the benign development of the token ecosystem has become an inevitable trend. Recent efforts have been conducted to analyze the token ecosystem of Ethereum across various dimensions. For example, \cite{victor2019measuring,chen2020traveling} analyze Ethereum-based ERC20 token networks from a graph perspective. Meanwhile,  studies~\cite{frowis2019detecting,chen2019tokenscope,di2021identification} attempt to detect inconsistent and abnormal behaviors in the ERC20 token ecosystem. Moreover, Fenu et al.~\cite{fenu2018ico} investigate the relationship between ICO and Ethereum contracts while \cite{conley2017blockchain,howell2020initial} summarizes the characteristics of successful tokens. However, none of these studies have explored the token ecosystem in EOSIO. The comparison study of the EOSIO token ecosystem and other blockchains (like Ethereum) can help to characterize different blockchains in terms of ICOs. Our paper is the first comprehensive work to study the EOSIO token ecosystem.

\subsection{Fake Detection}
The prosperity of blockchain systems and smart contracts also brings fraudulent activities. Fraudsters often make scams to defraud investors' assets. For example, some studies~\cite{chen2018detecting,BARTOLETTI2020259} show that Ponzi schemes with forged high-yield illusion were found in Ethereum to attract huge investments from victims. Similarly, many ICO parties also counterfeit fake users and fake transactions to make unreal prosperity of their ICO projects or DApps. Several recent studies have attempted to detect fake users and illegal activities. Farrugia et al.~\cite{farrugia2020detection} identify fake and illicit accounts over the Ethereum blockchain. Meanwhile, Huang et al.~\cite{huang2020characterizing} find some bot-like and malicious accounts in EOSIO while their study does not consider tokens of EOSIO. No previous work has identified fake tokens and fake users or transactions related to tokens. We propose an algorithm to detect fake tokens and recognize manipulation behaviors, thereby increasing investors’ vigilance against fake tokens and avoiding harmful investments.

\section{Conclusion and Future Work}\label{Conclusion}
To our best knowledge, we are the first to conduct a holistic measurement study on the EOSIO token ecosystem. After gathering a comprehensive dataset, we construct multiple graphs to characterize the tokens, token holders, and token creators, accompanied by a comparison study with Ethereum. We then analyze token transfer flows; this analysis also helps us to identify some abnormal trading patterns in EOSIO. Moreover, we propose an algorithm to detect tokens with fake users and fake transactions. Our study may help investors to be aware of abnormal behaviors of tokens to avoid harmful investments. This study offers many insightful findings, which help people have an in-depth understanding of the EOSIO token ecosystem and also raise many interesting open questions in this area: 1) Why have some inactive users created so many tokens with attempts to attack EOSIO? 2) What has occurred in the abnormal trading patterns? 3) What roles do the accounts play in each abnormal pattern? 4) Are there other relationships between the manipulators and fake accounts?

\bibliography{main.bib}

\clearpage
\end{sloppypar}
\end{document}